\documentclass[iicol,sn-mathphys-num]{sn-jnl}

\usepackage{graphicx}%
\usepackage{multirow}%
\usepackage{amsmath,amssymb,amsfonts}%
\usepackage{amsthm}%
\usepackage{mathrsfs}%
\usepackage[title]{appendix}%
\usepackage{xcolor}%
\usepackage{textcomp}%
\usepackage{manyfoot}%
\usepackage{booktabs}%
\usepackage{algorithm}%
\usepackage{algorithmicx}%
\usepackage{algpseudocode}%
\usepackage{listings}%

\usepackage{makecell}
\usepackage{url}
\usepackage{adjustbox}
\usepackage{caption}
\usepackage{subcaption}
\usepackage{soul}
\usepackage{natbib}
\usepackage{nomencl}
\makenomenclature

\theoremstyle{thmstyleone}%
\theoremstyle{thmstyletwo}%

\theoremstyle{thmstylethree}%

\begin{document}

\title[Article Title]{A Comparative Analysis of DNN-based White-Box Explainable AI Methods in Network Security}


\author[1]{\fnm{Osvaldo} \sur{Arreche}}\email{oarreche@purdue.edu}

\author*[2]{\fnm{Mustafa} \sur{Abdallah}}\email{abdalla0@purdue.edu}


\affil[1]{\orgdiv{Electrical and Computer Engineering Department}, \orgname{Purdue University in Indianapolis}, \orgaddress{\street{420 University Blvd}, \city{Indianapolis}, \postcode{46202}, \state{Indiana}, \country{USA}}}

\affil*[2]{\orgdiv{Computer and Information Technology Department}, \orgname{Purdue University in Indianapolis}, \orgaddress{\street{420 University Blvd}, \city{Indianapolis}, \postcode{46202}, \state{Indiana}, \country{USA}}}



\abstract{New research focuses on creating artificial intelligence (AI) solutions for network intrusion detection systems (NIDS), drawing its inspiration from the ever-growing number of intrusions on networked systems, increasing its complexity and intelligibility. Hence, the use of explainable AI (XAI) techniques in real-world intrusion detection systems comes from the requirement to comprehend and elucidate black-box AI models to security analysts. In an effort to meet such requirements, this paper focuses on applying and evaluating White-Box XAI techniques (particularly LRP, IG, and DeepLift) for NIDS via an end-to-end framework for neural network models, using three widely used network intrusion datasets (NSL-KDD, CICIDS-2017, and RoEduNet-SIMARGL2021), assessing its global and local scopes, and examining six distinct assessment measures (descriptive accuracy, sparsity, stability, robustness, efficiency, and completeness). We also compare the performance of white-box XAI methods with black-box XAI methods. The results show that using White-box XAI techniques scores high in robustness and completeness, which are crucial metrics for IDS. Moreover, the source codes for the programs developed for our XAI evaluation framework are available to be improved and used by the research community.
}

\keywords{Explainable AI, XAI Evaluation, Intrusion Detection Systems, LRP, Integrated Gradients,   Network Security,  DeepLift, White-box AI, NSL-KDD, CICIDS-2017, RoEduNet-SIMARGL2021}



\maketitle

\section{Introduction}\label{sec:introduction}

The development of AI brought the advancements of automation to intrusion detection systems (IDS)~\cite{buczak2015survey,dina2021intrusion}. A few examples of such endeavor are decision trees (DT)~\cite{ferrag2020rdtids,al2021intelligent}, random forests (RF), support vector machines (SVM), and naive bayes (NB)~\cite{amor2004naive,panigrahi2022intrusion}.
Except for decision trees for basic IDS, the security analyst in charge often lacks explanations about the AI's decisions~\cite{arisdakessian2022survey,sabev2020integrated}. More specifically, these earlier studies using AI focused more on the accuracy of categorization of various AI systems without providing insight into their reasoning or behavior. This limitation raises the acute need to take advantage of the new field of explainable AI (XAI) to enhance the comprehensibility of AI decisions in IDS~\cite{das2020opportunities}. \sloppy

Several recent studies (e.g.,~\cite{mahbooba2021explainable,patil2022explainable,islam2019domain,roponena2022towards,owad}) have begun investigating the use of XAI for IDS. 
For instance, a human-in-the-loop method to intelligent IDS was offered in an abstract design by the work~\cite{roponena2022towards}. In another example, the NSL-KDD benchmark dataset~\cite{dhanabal2015study} was the only tool utilized by the work~\cite{mahbooba2021explainable} to extract decision rules, using simply a straightforward decision tree algorithm. Differently,
Local interpretable model-agnostic explanations (LIME)~\cite{dieber2020model} were included by the work~\cite{patil2022explainable} to only give local explanations for a single AI model (SVM) for the utilized IDS. In a distinct framing, The CIA principle ({{``C'' for confidentiality of information, ``{I'' for the integrity of data, and ``{A'' for availability) was applied in the work~\cite{islam2019domain} to improve the explainability of the AI model on the CICIDS-2017 benchmark intrusion dataset. In essence, these efforts lack a comprehensive metric evaluation of XAI techniques in a diverse set of intrusion datasets.\vspace{-0.2mm}

The main challenge of implementing XAI tools into network intrusion detection is validating such tools, testing their quality, and evaluating the pertinent security metrics. This step is paramount to build confidence in applying them in practical settings, considering that they will eventually be widely spread in network security to help security analysts be more effective and accurate in their tasks.

Diving deeper into XAI, a concept adopted in this work presented on~\cite{evaluating6metrics} that divides XAI into black-box and white-box methods. The first refers to the methods that do not have access to the internal parameters and architecture of the AI models. Therefore, black-box XAI methods rely solely on approximating the model's behavior by observing the relationship between input and output. These characteristics make the black-box XAI methods versatile and agnostic, which applies in several AI models (e.g., LIME~\cite{LIME} and SHAP~\cite{SHAP}. 
In contrast, white-box XAI methods (i.e., the focus of this current study) are deemed transparent because of their access to the model's parameters and architecture. The white-box methods use the knowledge and access to the model to generate explanations, which is advantageous because it is not an approximation but the model itself. This difference makes them more reliable and accurate, providing more insights when reasoning their decisions, with the drawback of being model-specific. The white-box methods analyzed in this article are used in tandem with neural networks,  including Integrated Gradients (IG)~\cite{IG}, Layer-wise Relevance Propagation (LRP)~\cite{LRP}, and DeepLift~\cite{DeepLift,DeepLift2,DeepLift3}. 

This research intends to evaluate white-box XAI techniques for network intrusion detection systems by proposing a new XAI framework based on the previous works~\cite{evaluating6metrics,osvaldo}. The metrics proposed by~\cite{evaluating6metrics} are modified for network intrusion detection systems and used to assess XAI algorithms. The measures in question include completeness, efficiency, stability, robustness, descriptive accuracy, and sparsity. Here is a definition of these different metrics (or measurements).

\begin{itemize}
\item \textbf{Descriptive Accuracy:} This metric disregards the top features when training and testing the AI model to evaluate if the disregarded feature affected the model's accuracy. For example, if a feature that detects intrusions is significant, removing that feature should reduce the accuracy of the AI model's intrusion prediction. Hence, a significant accuracy loss indicates that such features have a high XAI explainability power.

\item \textbf{Sparsity:} The sparsity gauges the relationship between the features' significance scores, which is quantified by comparing the number of features that fall below a threshold. For instance, if eight out of 10 features fall below a small threshold (around zero), it indicates that two features have a strong impact on the model's judgment according to the scores assigned by the XAI, indicating a high level of sparsity. This could help security analysts to narrow their search when monitoring network traffic.

\item \textbf{Efficiency:} The time it takes for an XAI method to generate an explanation is used to measure its efficiency. Since it is more practical to have explanations created rapidly rather than slowly, this statistic is significant since it assesses how well the XAI technique applies to real-world systems. Since assisting security analysts is the ultimate objective, the system should be able to produce precise XAI explanations quickly. 

\item \textbf{Stability:} The XAI method's consistency in producing explanations is gauged by the stability metric, which is quantified by examining the number of overlapped top features between several running experiments conducted under identical circumstances (i.e., to check if the produced explanation by the XAI method is the same for the same sample). An XAI approach with higher stability can be trusted more by the security analyst.

\item \textbf{Robustness:} 
The capacity of an XAI approach to producing the same explanations even when the incursion traits are perturbed is known as its robustness, even when an adversarial assault or computational faults may be the cause of this disturbance. In this experiment, the adversarial model from the work~\cite{adversary} is modified to work with the IDS datasets considered in our current work. In short, to create this adversarial model, one highly biased model is trained by being heavily biased in one feature for prediction, and another model with all the features plus a new feature designed to perturb (i.e., with random values) the XAI explanation technique is trained.
The XAI approach under test will produce explanations based on these two trained models. The attack consists of displaying the adversarial explanation to the security analyst. For example, this might compromise the framework's integrity and pass off an attack as regular network traffic by severing the explanation from the underlying behavior.

\item \textbf{Completeness:} 
The capacity of an XAI approach to correctly explain every potential network traffic sample, including corner instances, is referred to as its completeness feature. An incomplete XAI approach makes it easier for hackers to take advantage of and fool the system into producing corrupted output. Another important distinction about the nature of Completeness when comparing white-box and black-box XAI methods is the inherent completeness of white-box methods ~\cite{evaluating6metrics} due to its access to the model's innards. Its access allows the white-box XAI to fulfill the requirement of valid explanations without testing all the samples individually (i.e., black-box XAI methods need to do this to check their completeness) since the model is known and it is taken into consideration. Hence, for a white-box method, a wrong explanation would imply something inappropriate with the model, while the same case cannot be determined for a black-box XAI case (i.e., it could be the model or the XAI approximation method).

We stress that an XAI approach automatically becomes more resilient if it is complete (i.e., it can determine whether the explanation is valid, increasing the end user's trust in it). In summary, the robustness metric in this study measures how resilient an XAI system is to an adversarial assault, while the completeness metric verifies that each sample has a legitimate explanation.

\end{itemize}

We investigate these aforementioned six assessment measures on three widely used white-box XAI techniques, which are IG~\cite{IG}, DeepLift~\cite{DeepLift,DeepLift2,DeepLift3}, and LRP~\cite{LRP}.

LRP~\cite{LRP} works by assigning relevance scores to each input feature or a neuron which indicates its contribution of the predictions of deep neural networks. The goal of LRP is to assign relevance scores to each input feature or neuron in the network by recursively using back-propagation, indicating its contribution to the output prediction. Such contribution obeys propagation rules guaranteeing the sum of the scores are conserved~\cite{LRP}. Similarly, DeepLift follows the same conservation principle of LRP, but it enhances the method by adding a new axiom, which dictates how to distribute the relevance scores (i.e., feature importance scores)~\cite{DeepLift,DeepLift2,DeepLift3}. 

Integrated Gradients (IG) works by obeying two principles. Sensitivity and Implementation Invariance. The first takes into consideration the contribution of each feature to the outcome, in other words, if such a feature contributes to the outcome it should have a non-zero contribution. The second principle, Implementation Invariance, premises the attribution method should not take into account models' particularities, meaning that two different network models that yield the same result considering the same input should have the same contributions. Consequentially, it achieves both principles by taking into consideration the model's gradients considering each input feature and multiplying them by the difference between the resulting output and a baseline~\cite{IG}.

Furthermore, we detail the sequential steps taken to construct our white-box XAI assessment measures and their outcomes for each of our six evaluation metrics (i.e., descriptive accuracy, sparsity, efficiency, stability, completeness, and robustness), utilizing three distinct network intrusion datasets. The first dataset was gathered from the SIMARGL project (funded by the European Union) and is the most recent RoEduNet-SIMARGL2021 dataset~\cite{mihailescu2021proposition}. The dataset is invaluable for network intrusion detection systems since it includes characteristics computed from live traffic in addition to realistic network traffic. The second dataset, CICIDS-2017~\cite{panigrahi2018detailed}, is a benchmark intrusion detection dataset with varying attack patterns that was developed in 2017 by the University of Brunswick's Canadian Institute for Cybersecurity. The NSL-KDD dataset~\cite{dhanabal2015study} is the last dataset and is well-known for network intrusion detection systems widely applied as a benchmark. 

We assess the three white-box XAI techniques (LRP, IG, and DeepLift) for the DNN model for each dataset, and we discuss the findings of the six assessment metrics produced for each XAI technique, by using assessment and comparison metrics to validate these XAI approaches, our study improves in the use of white-box XAI methods for network intrusion detection systems. These metrics cover AI model qualities such as descriptive accuracy, sparsity, and explainability, as well as network security needs like stability, robustness, and efficiency. As a result, our framework enables the integration of XAI into network intrusion detection systems, opening the door to increased  development in this critical field of study.

\noindent \textbf{Compendium of Contributions:}
Below is a summary of our principal contributions.

\begin{itemize}
    \item We propose a comprehensive approach for assessing local and global white-box XAI methods in the context of network intrusion detection. 

    \item We examine six distinct assessment measures for DeepLift, IG, and LRP, three well-known white-box XAI approaches.

    \item We test our XAI assessment framework on Tensorflow-based Deep Neural Network AI models and three network intrusion datasets.

    \item We make our source codes available so the community may use them for XAI assessment frameworks for network intrusion detection and expand upon them with additional datasets and models.\footnote{The URL for our  source codes is:
\url{https://github.com/ogarreche/XAI_Whitebox}}
\end{itemize}


\textbf{Paper Organization:} The paper is structured as follows.

\begin{itemize}
    \item Section~\ref{sec:introduction} introduces the need for White-box XAI methods to explain black-box AI models in the context of Network Intrusion Detection Systems (NIDS).

    \item  Section~\ref{sec:related_work} gathers related works and current efforts in the field to address the need briefly explored in the Introduction.

    \item  Section~\ref{sec:background} discusses further the problem statement details and the preliminaries before introducing the proposed framework.

    \item Section~\ref{sec:framwork} presents the main components of the proposed white-box XAI framework. 

    \item Section~\ref{sec:evaluation} presents the foundations of evaluations of the proposed framework applied to the network intrusion datasets.

    \item Section~\ref{sec:metric_analysis} analyzes the results of the proposed XAI metrics in depth.

    \item  Section~\ref{sec: Discussion} discusses the limitations, insights, and future directions of the generated results. 

    \item Section~\ref{sec: conclusion} presents a summary of the study, the main insights, and the concluding remarks.

\end{itemize}

\definecolor{osvaldo}{RGB}{219, 48, 122}



\section{Related Work}\label{sec:related_work}

\begin{table*}[t]
\caption{A comparison between different aspects of our own work and prior relevant works on XAI for network intrusion detection (including methods for explanations, datasets, AI models, XAI evaluation, and application domain).} 
\resizebox{\textwidth}{!}{
\begin{tabular}{c|c|c|c|c|c|c}
\textbf{Paper} & \textbf{Dataset}  & \textbf{Model} & \textbf{Explainer} & \textbf{Evaluating XAI} & \textbf{White-box XAI} & \textbf{IDS} 
\\
\hline
\hline
Our  Work  & \makecell{CICIDS-2017, NSL-KDD,  \\ RoEduNet-SIMARGL2021} & \makecell{DNN}   & DeepLift, IG, LRP & Yes & Yes & Yes\\
\hline

\makecell{E-XAI: Evaluating Black-Box \\ Explainable AI
~\cite{osvaldo}} & \makecell{CICIDS-2017, NSL-KDD,  \\ RoEduNet-SIMARGL2021} & \makecell{ADA, KNN, MLP \\ DNN, RF, SVM, LGBM}   & SHAP, LIME & Yes & No & Yes\\
\hline

Evaluating XAI
in Security~\cite{evaluating6metrics} & \makecell{Malware Genome Project,\\CWE-119} & CNN, MLP, RNN & \makecell{SHAP, LIME , LEMNA,\\ IG, LRP, Gradients} & Yes  & Yes & No\\
\hline

DeepAID~\cite{deepaid} & \makecell{Kitsune-Mirai, HDFS, \\ CICIDS-2017, LANL-CMSCSE }  & \makecell{DNN,   LSTM, GNN \\ Autoencoders, Kitsune } & GLGV,   DeepLog,   DeepAID & No  & Yes & Yes\\
\hline

Kitsune~\cite{kitsune} & \makecell{Kitsune   (OS Scan, \\Fuzzing, and Mirai)}  & \makecell{Kitnet, GMM, SVM,\\ DNN, Autoencoders}  & Kitsune & No  & No & Yes\\
\hline
OWAD~\cite{owad}  & \makecell{Kyoto 2006+, BGL, \\ LANL-CMSCSE}  & \makecell{DNN, Autoencoder, DeepLog, \\ APT (GLGV), GAN, LSTM} & OWAD, CADE, TRANS & No  & No & Yes\\
\hline

\makecell{Feature-oriented \\  Design~\cite{Feature_oriented_Design}} & NSL-KDD  & CNN,   DNN  & \makecell{LIME, and Saliency View} & No & Yes & Yes\\

\hline
\makecell{Explainable ML \\ Framework~\cite{Explainable_Machine_Learning_Framework}} & NSL-KDD &\makecell{ DNN,   RF, SVM, KNN,\\ ResNet50}  & SHAP & No   & No & Yes\\

\hline
\makecell{Why should I trust \\ your IDS?~\cite{Why_should_I_trust_your_IDS,Novel_IoT_Based}} & NSL-KDD, UNSW-NB15  & DNN &  SHAP, LIME, RuleFit & No  & No & Yes\\
\hline

Fooling LIME and SHAP~\cite{adversary} &  \makecell{COMPAS,  German Credit, \\ Communities and Crime}  & RF &  SHAP, LIME& No  & No & No\\
\hline

\end{tabular}
}
\label{tbl:related_works_XAIEval}
\end{table*}



\textbf{Existing Efforts in Leveraging XAI for IDS:}
The survey~\cite{survey} provides an overview of current attempts to create a framework for developing Explainable AI-based Intrusion Detection Systems (XAI-IDS) and proposes a standard taxonomy. It discusses the differences between opaque black-box and transparent white-box models, examining the trade-offs between performance and explainability. In AI, black-box models (e.g., Neural Networks, and Random Forest) are difficult to interpret, while white-box AI models (e.g., Logistic Regression) are easier to explain but often perform poorly. Additionally, there are black-box and white-box XAI techniques~\cite{evaluating6metrics}. Black-box XAI methods, like SHAP and LIME, do not access the model's parameters or architecture, whereas white-box methods, such as LRP, IG, and DeepLift, do. \sloppy

Furthermore, surveys~\cite{white1,white3,white5} gather scattered data and improvements across different proposals for explaining black-box AI models. It can be divided into two themes of general challenges and future directions, for instance taking into consideration the user experience, formalism, or model assumptions, and research directions/challenges of XAI based on AI life cycle's phases (e.g., specific situations that might occur into the conception, developing or deploying phase). Also, this work analyzes its application in the medical area and proposes a further investigation into other areas such as IoT. In the same direction, ~\cite{white2} analyzes XAI to make them more understandable and transparent, proposing a taxonomy of transparent models and post-hoc techniques for deep learning models. Moreover, it discusses the need for standardized metrics, addressing issues related to robustness, confidentiality, accountability, privacy, and its evolving regulatory demands. Plus, the DARPA XAI program~\cite{white4} focuses on evaluating new techniques and exploring modified deep neural networks, causal models, and induction techniques with the goal of creating interpretable AI models and navigating the trade-off between explainability and performance. Lastly, the work~\cite{white6} focuses on XAI medical application, it explores an in-depth explanation considering different works of machine learning models in the field considering accountability and ethical considerations. It also explores that the explanations can be tampered leading to unreliable explanations.

Some recently published papers initiated the application and analysis for XAI in the field of IDS~\cite{mahbooba2021explainable,patil2022explainable,islam2019domain,roponena2022towards}.
In general, the survey~\cite{survey} proposes a new XAI framework for IDS and gathers most of the studies related to the field. It is divided into phases (i.e., pre-processing, modeling, and explainability). The first converts the input data into better quality, the second generates the explanations, and the later customizes the dashboard accordingly to the target user. Other examples include~\cite{islam2019domain}, which proposed an explainable AI model using CICIDS-2017 in tandem with the CIA principles, ~\cite{mahbooba2021explainable} used DT to understand the IDS rules applied to the NSL-KDD dataset, and the work~\cite{patil2022explainable} which provided explanations locally using LIME combined with SVM in IDS. 

Plus, the works~\cite{Why_should_I_trust_your_IDS, Novel_IoT_Based} use a DNN model in the context of anomalous traffic (i.e, normal or suspicious traffic) in tandem with NSL-KDD and UNSW-NB15 datasets to propose an XAI framework focused on IoT. It leverages the explanations from RuleFit, SHAP, and LIME, to achieve trust while generating the reason behind black-box AI decisions in order to aid security users. In a different light, the research in this paper proposes an XAI framework that uses DNN models, for the multiclassification case (i.e., different attacks). Such framework employs IG, LRP, and DeepLift to explain black-box AI to security analysts, and it evaluates the explanation quality in the light of six novel metrics~\cite{evaluating6metrics}.   

There are also other significant works in the field. ~\cite{owad} developed a framework called Open World Anomaly Detection (OWAD) for dealing with drifting in deep learning models. Meanwhile, DeepAID~\cite{deepaid} tackled some shortcomings in unsupervised Deep Learning binary detection systems. Kitsune~\cite{kitsune} developed an anomaly detection tool for camera network systems. In summary, these mentioned works are in  Table~\ref{tbl:related_works_XAIEval}, which compares them with this current work.

\textbf{XAI Evaluation for Network Intrusion Detection:} In the literature, there is a gap in applying and evaluating XAI techniques in network intrusion detection. For instance, the papers~\cite{evaluating6metrics,adversary,osvaldo} proposed and evaluated XAI techniques in limited scenarios. First, the article~\cite{evaluating6metrics} innovated in laying six new metrics for XAI in deep learning applied to security, it analyzes white-box and black-box XAI methods considering the CWE-119 and Malware Genome Project datasets. Nonetheless, they focused on other security cases from those considered in our current work. CWE-119 is a dataset focused on uncovering vulnerabilities for implementing code, while the other dataset, is a compendium of malware samples that target Android-based applications.

The paper~\cite{adversary} developed an adversarial attack on LIME and SHAP to test their robustness. It trained two models—one extremely biased and the other innocuous with an unrelated feature—to deceive LIME and SHAP into creating seemingly correct explanations for attack samples. This study used datasets outside the security domain: COMPAS (criminal history, prison time, and demographic attributes), Communities and Crime (socioeconomic and crime information), and German Credit (financial and demographic information of loan applicants). The authors shared their code publicly, which we adapted for testing the XAI method's robustness in our network intrusion detection context for our different white-box XAI methods and datasets.

The paper~\cite{osvaldo} proposed and applied the XAI evaluation metrics (i.e., Descriptive Accuracy, Sparsity, Efficiency, Stability, Completeness, and Robustness) to black-box XAI techniques (i.e., LIME and SHAP). According to~\cite{evaluating6metrics}, black-box methods infer AI model behavior from the input $x$ and output $f(x)$ without knowledge of model parameters or architecture. This method is useful for auditing or experimenting with systems without model access, but it lacks valuable model information. Conversely, white-box methods, considered in this current work, have access to model architecture and parameters, avoiding the drawbacks of black-box methods but requiring prior model access and being tied to specific models (e.g., IG and LRP for Neural Networks). SHAP, however, is model-agnostic.

\textbf{Contribution of This Work:} Differently from~\cite{evaluating6metrics,adversary,osvaldo} and the works from Table~\ref{tbl:related_works_XAIEval}, this work introduces a framework for evaluating XAI targeted for the network intrusion detection systems context, using White-box XAI methods. The framework is applied to RoEduNet-SIMARGL2021,CICIDS-2017, and NSL-KDD, invaluable datasets, and it is mainly focused on using DNN models. Moreover, the six XAI metrics (i.e., Descriptive Accuracy, Sparsity, Efficiency, Stability, Completeness, and Robustness) are considered. These metrics are used with white-box XAI methods (LRP, IG, and DeepLift), evaluating the most crucial features selected by these techniques. It is invaluable to mention the increased Robustness~\cite{adversary} of these white-box XAI techniques. They show superior performance when compared to their counterparts~\cite{osvaldo}. Although their inherent nature of using the model's knowledge makes them complete in the Completeness sense~\cite{evaluating6metrics}, this work takes the extra effort and runs the completeness experiment either way in, a reduced form, to display its superiority when compared to black-box XAI experiment baseline in~\cite{osvaldo}. The efforts in this work represent an invaluable step in the direction of deploying white-box XAI in network IDS.


\section{The Problem Statement}\label{sec:background}

This section lays the foundation necessary for this research, before discussing its impacts. The topics are the shortcomings of black-box AI, hence the reason for the need for XAI, and lastly the evaluation of such methods applied to network IDS together with its challenges.
\vspace{-2mm}

\subsection{Network Intrusion Types}

Network intrusion attacks come in several types. This study uses the MITRE ATT\&CK framework~\cite{strom2018mitre}, dividing the intrusions in:

\textbf{[MITRE ATT\&CK ID: T1595] Network Scanning / Probe Attack:} This class of assaults comprehends probing and gathering information for a network in order to find vulnerable spots~\cite{chen2022intrusion}. However, not all probe attacks are considered port scan attacks. For instance, ping sweeps, and DNS zone transfers are other methods in this category~\cite{gorodetski2002attacks,skwarek2019characterizing}.

\textbf{[MITRE ATT\&CK ID: TA0001, T1110, T1078] Remote to Local Attack (R2L):} This intrusion comprehend initial access. In other words, an attacker without permission on a network acquires access to a machine or network. From there, the agent can acquire more privileges to expand its access over the network or systems, in other words, U2R.

\textbf{[MITRE ATT\&CK ID: TA0004, T1078] User to Root Attack (U2R):} This consists of an agent exploiting the machine accesses until it achieves control of the root. Such an attack jeopardizes the entire system since the attacker would have administrator privileges. Nonetheless, the agent would still need to perform R2L to gain its initial access.

\textbf{[MITRE ATT\&CK ID: T1046] Network Service Discovery / PortScan (PS):} This category is about an agent surveying spots that can be vulnerable to attacks in the targeted network. Its modus operandi is connecting to various ports without ever finalizing the communication channel. Then, it uses its acquired knowledge to create a mapping of the network to potential vectors of attack~\cite{lee2003detection}.

\textbf{[MITRE ATT\&CK ID: T1498] Denial of Service (DoS) / Network Denial of Service:} In this case the agent overwhelms the victim with requests to connect with the aim of making the network unavailable. Such attacks usually consume the victim's memory rendering it temporarily unavailable~\cite{CICIDS,SENSOR}.

\textbf{[MITRE ATT\&CK ID: T1110] Brute Force:} This assault is a direct approach to guessing the password of the target. It usually uses a list of commonly used passwords~\cite{CICIDS}.

\textbf{[MITRE ATT\&CK ID: TA0001, T1659, T1189] Web Attack / Initial Access:} The category embodies gaining access to a network or container by exploiting web weak spots. For instance, a malicious agent could use a bug/glitch or an overlooked configuration to take advantage of a public application. Another example includes Drive-by 
Compromise~\cite{Web_attack_Ref2}, which, usually, does not serve as a remote server access entry point~\cite{strom2018mitre}.

\textbf{[MITRE ATT\&CK ID: TA0001] Initial Access / Infiltration:} It includes the use of techniques (e.g., spear-phishing or web-server weakness exploitation) to obtain access to a network. It can vary from a password swap to the capture of a valid account or other services.

\textbf{[MITRE ATT\&CK ID: T1584.005, T1059, T1036, T1070] Botnet / Compromise Infrastructure :} It comprises, often, remote attacks that are executed by compromised devices using bots or scripts that could mimic the behavior of a human. It is a powerful, common, and scale-able technique that can aim for different attack vectors at the same time~\cite{stone2009your}.

\textbf{Regular traffic:} It comprises all other traffic that is not an attack, therefore regular or benign network traffic.

\subsection{Intrusion Detection Systems}

The malicious agents' arms race against cyber-security sprouted evolving threats to invaluable infrastructure~\cite{khan2021m2mon,hussain2021noncompliance}. As a response, IDS became increasingly critical to containing and avoiding these assaults (e.g., internal and external) against networks~\cite{mirzaei2021scrutinizer}. In a traditional sense, IDS assumes a signature from attacks, and if such a pattern matches it flags it as suspicious traffic to be investigated~\cite{lukacs2015strongly}. In addition, with the evolving AI over the last years, it became the norm to incorporate AI into the IDS to enhance its capabilities~\cite{kim2020ai}.

\subsection{Black-box AI Models and its caveats}

The aforementioned arms race increased the complexity of the competition, and now complex, yet efficient, black-box AI models are employed to defend and flag suspicious traffic. Nonetheless, a side-effect of black-box AI models is to humanly understand it and to extract how it concluded because of the complicated relations it can abstract from features and scenarios. This challenge is a common black-box issue across many models (e.g., DNN, RF, and others) and it is aggravated when they predict an outcome that is not correct. This becomes a sensitive issue since network security is heavily based on trust in the system. Hence, it is paramount to provide valuable explanations behind the AI rationale for IDS~\cite{botacin2021challenges}.

\subsection{Main Categories of Explainable AI (XAI)}

XAI is a new field of research created to tackle the explainability issue behind complex AI models. It embodies analyzing the features and other attributes to understand the contribution of a sample responsible for an outcome, either a regression or classification. Usually, XAI is categorized into the following: (1) XAI methods able to explain local instances (i.e., one sample), global instances (i.e., many samples), or both. (2) model-agnostic techniques (that can explain any AI model) or model-specific techniques (tailored for specific AI models). According to~\cite{evaluating6metrics}, black-box methods infer AI model behavior from the input $x$ and output $f(x)$ without knowledge of model parameters or architecture. This method is useful for auditing or experimenting with systems without model access, but it lacks valuable model information. Conversely, white-box methods have access to model architecture and parameters, avoiding the drawbacks of black-box methods but requiring prior model access and being tied to specific models (e.g., IG and LRP for Neural Networks). SHAP, however, is model-agnostic. (3) XAI methods can also be divided by Post-hoc and Ante-hoc (i.e., intrinsic), according to~\cite{antehoc}. Ante-hoc methods provide explanations directly from the model without additional steps after training. For example, small Decision Trees are easy to understand and explain without any further techniques besides analyzing the tree and its rules. In contrast, post-hoc XAI methods, which are the focus of this article, are applied after training to extract further understanding and explainability from black-box AI models~\cite{antehoc}. 

\subsection{Benefits of XAI for Network IDS}

The lack of interpretability scenario is a challenge for human analysts, as usually, they might analyze a long number of logs to explain suspicious behaviors. Besides, a common scenario is the inability to exactly where the AI model can be improved. Plus, the situation is aggravated in the realm of NIDS, since both aspects are invaluable (i.e., clear explanation and high accuracy) to build trust in the system's capabilities. Therefore, XAI systems or frameworks contribute with a crucial aspect to enable the clarification of the AI rationale, which facilitates the investigations, making the analysts more efficient.

\subsection{Challenges of XAI for IDS and Need for Evaluating XAI}

There are challenges to applying XAI to NIDS. Accordingly, to the metrics cited in~\cite{osvaldo,evaluating6metrics}, the white-box and black-box XAI techniques do not fulfill all the requirements to be employed in an IDS system, they often lack performance in some key areas. Nonetheless, Section~\ref{sec:metric_analysis} analyses the aforementioned metrics, and it demonstrates that white-box XAI methods are complete by default and more robust than black-box methods. Next is a summary of the challenges:

\begin{itemize}

   \item \textbf{Not Transparent:} Often, the IDS have black-box AI models, which brings the challenge of understanding and explaining them. Consequentially, it impairs the work of security analysts when they proceed with auditory actions and even guard their system against threats.

  \item  \textbf{XAI Application in Network Security is still Limited:} Network intrusion detection poses unique challenges for XAI~\cite{https://doi.org/10.48550/arxiv.1812.07858}, which entails a need to improve their interpretation techniques for the field. Differently from other areas (e.g., Computer vision and text analysis), XAI is not yet fully consolidated in NIDS. 

    \item    \textbf{Comparison and Evaluation:} Since XAI in NIDS is not yet consolidated, it brings a need for criteria and metrics to evaluate them~\cite{capuano2022explainable}. These need to embody the requirements of securing networks, including efficiency, robustness, stability, and reliability while keeping the requirements for AI properties, which are transparency, accuracy, and explainability.

  \item  \textbf{Robustness and Accuracy:} XAI methods for NIDS need to display valid and robust explanations, while being efficient and accurate at the same time. These conditions produce the reliability and trust of XAI methods, which ensures their application in NIDS.
\end{itemize}

Having defined prior related works and main challenges that motivate the need of evaluating XAI, we next detail our proposed evaluation framework of white-box XAI methods.

\section{Framework}\label{sec:framwork}

This paper aims to evaluate a newly white-box XAI framework applied to NIDS using the proposed XAI metrics. Besides, this work contributes to advancing XAI into NIDS, helping security users obtain in-depth abstractions of network traffic by extracting the white-box XAI evaluation and enhancing their trust through the evaluated metrics. Consequentially, it would augment their capability to avoid different network attacks and suspicious scenarios. The framework can be divided as follows.

\begin{figure}[t]
   \centering
\includegraphics[width=\linewidth]{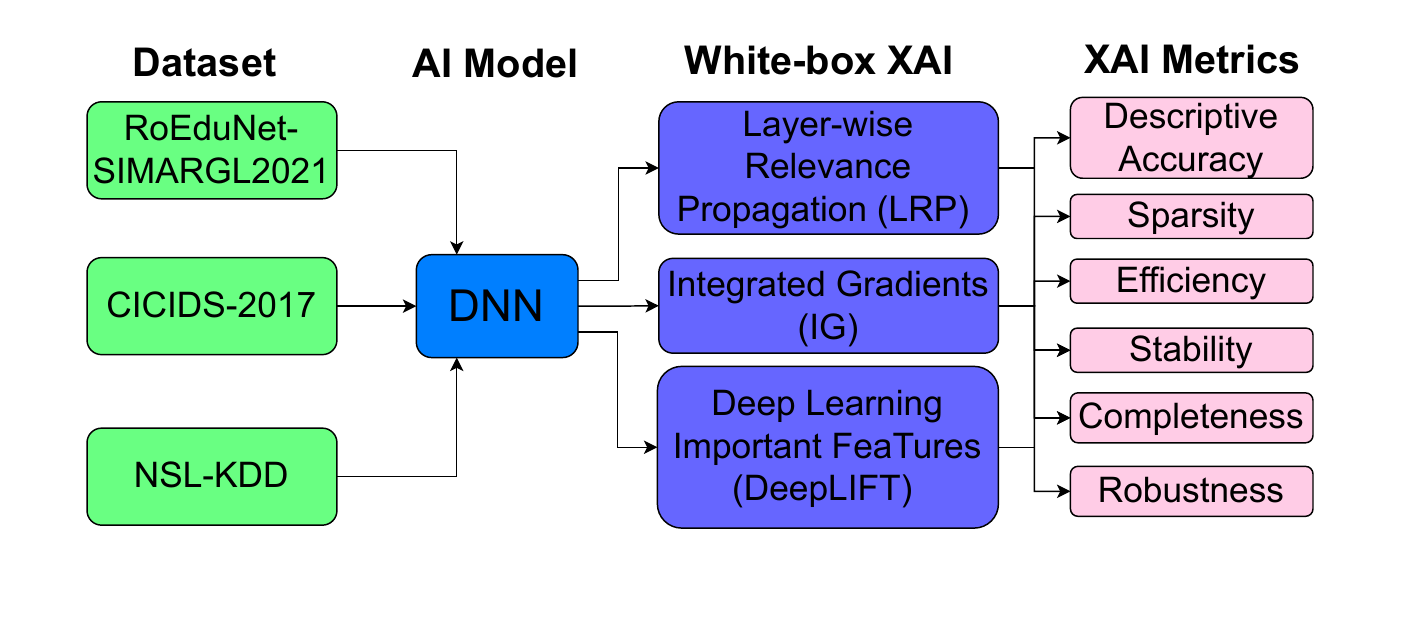}
\vspace*{-6mm}
   \caption{A diagram of the XAI framework for evaluation of network intrusion detection. It considers six evaluation metrics, three white-box XAI methods, a neural network AI model, and three invaluable intrusion datasets.} 
   \label{fig:Flow_code}
   \vspace{-5mm}
\end{figure}

\subsection{Overview of the XAI Evaluation Framework}

The proposed framework starts by preparing the invaluable NIDS datasets (i.e., CICIDS-2017, NSL-KDD, RoEduNet-SIMARGL2021). Next, after the data is properly loaded and pre-processed, it is forwarded to the respective DNN model to train and test over several intrusions (i.e., Multi-class classification problem). The classes include Port Scanning, Brute Force, R2L, and others. Later, the neural network itself and predictions are moved towards the white-box XAI methods, namely LRP, IG, and DeepLift to generate the desired explanations. The last part includes evaluating them in the optics of the evaluation metrics (i.e., Robustness, Completeness, Stability, Sparsity, Descriptive Accuracy, and Efficiency).

\subsection{In-Depth XAI Evaluation Pipeline Components}

We now detail the low-level components of our XAI pipeline, as shown in Figure~\ref{fig:Flow_code}.

\textbf{Load the Network Intrusion Datasets:} The first step is to load the network intrusion datasets, namely  NSL-KDD~\cite{dhanabal2015study}, CICIDS-2017~\cite{panigrahi2018detailed}, and RoEduNet-SIMARGL2021~\cite{mihailescu2021proposition}.

\textbf{Black-box AI Models:} After preprocessing the dataset, we train black-box AI models, splitting the data into 70\% for training and 30\% for testing. We use deep neural networks (DNN) and choose specific parameters for each model to ensure optimal performance (see Section~\ref{sec:evaluation}).

\textbf{AI Models Selection Criteria:} The DNN models used in this work were chosen because they are widely used in similar IDS studies (e.g.,~\cite{SENSOR,CICIDS,dhanabal2015study}), and to examine the impact of XAI-based feature selection on AI model performance in IDS. Using white-box methods like IG, LRP, and DeepLift necessitates working with neural networks to maintain consistency with IDS literature when comparing different models paired with XAI techniques.

\textbf{XAI Methods:} After constructing the black-box DNN model, we explain these models using three popular white-box XAI methods: LRP~\cite{LRP}, IG~\cite{IG}, and DeepLift~\cite{DeepLift}. These methods evaluate black-box AI models both globally and locally and are post-hoc, meaning they are applied after the AI model is trained and its predictions are generated. These model-dependent methods are well-known for evaluating black-box neural network models~\cite{evaluating6metrics}.

\textbf{Rationale for Choosing IG, LRP, and DeepLift:} We selected these three white-box XAI methods for their popularity, characteristics, and ease of implementation~\cite{evaluating6metrics,innvestigate}. Given the rising demand for neural network models to tackle complex problems, explaining such results is crucial, making IG, LRP, and DeepLift compelling choices since they are model-specific. Moreover, these approaches tend to be more robust, which is essential for IDS applications. Section~\ref{sec:evaluation} demonstrates that these techniques generally outperform black-box XAI methods. Additionally, these methods are accessible due to their available libraries for experimentation~\cite{innvestigate}.

The proposed pipeline assesses the three XAI techniques utilizing the following metrics: robustness, sparsity, efficiency, completeness, descriptive accuracy, and stability. Descriptive accuracy checks if the accuracy of an AI model drops after the most significant features, according to the respective XAI technique, are not present. Sparsity checks if the explanation is distributed among many features or concentrated on a few. Stability assesses the consistency of an XAI explanation across multiple trials under the same conditions. Efficiency measures the time taken by the XAI method to produce explanations. Robustness tests the XAI method's resistance to malicious actors trying to disguise attacks. Completeness evaluates whether the XAI method can provide correct explanations for every instance, including corner cases.

\subsection{Step-by-step Algorithms for Generating XAI Evaluation Metrics}

We detail the step-by-step process for generating our six XAI evaluation metrics: descriptive accuracy, sparsity, efficiency, stability, robustness, and completeness.\vspace{2mm}

\noindent \textbf{C.1.  Descriptive  Accuracy Step-by-Step{\hrule}}
\hfill \break
\textbf{Prerequisite:} Feature importance list from IG, LRP, DeepLift.

For each XAI method for DNN and each of the three datasets (NSL-KDD, CICIDS-2017, and RoEduNet-SIMARGL2021):
\vspace{-1mm}
    \begin{enumerate}
        \item Define k = (0, 5, 10, 25, 50, 70), the number of features to be removed in importance order.
        
        \item For each k value, obtain the AI model's accuracy. For k = 0, the model uses all features. For k = 5, the model uses all features except the top 5.
        
        \item Plot k on the X-axis and the AI model's accuracy on the Y-axis.
        
        \item Plot and analyze the Descriptive Accuracy graph using these axes. \hrule
        \vspace{1mm}
    \end{enumerate}


\noindent \textbf{C.2. Sparsity Step-by-Step\hrule}
\hfill \break
\textbf{Prerequisite:} Feature Scores obtained with DeepLift, IG, or LRP. Feature scores are normalized to the range [0,1].

For each of the three datasets (RoEduNet-SIMARGL2021, CICIDS-2017, and NSL-KDD), the Sparsity metric is applied as hereby:
\vspace{-1mm}
\begin{enumerate}
    \item Create an X-Axis with values [0.0, 0.1, 0.2, 0.3, 0.4, 0.5, 0.6, 0.7, 0.8, 0.9, 1.0] referring to the threshold compared to the feature scores.

    \item For each X-axis value (the threshold), apply the formula in Equation~\eqref{eq:sparsity_score} to determine the sparsity score.
    \begin{equation}\label{eq:sparsity_score}
\resizebox{0.7\columnwidth}{!}{$Sparsity = \dfrac{No\_Features\text{ s.t. } \{Feature\_Score \le \tau\}}{Total\_Number\_Features}.$}   
\end{equation}

    
\item Each result produced shapes the Y-axis, referring to the Sparsity score for each X-axis value. The Y-axis is normalized by dividing its values by the total number of features.

    \item Use the Y-axis and X-axis to plot the Sparsity graph.

    \hrule
    \vspace{2mm}
\end{enumerate}

\noindent \textbf{C.3. Efficiency Step-by-Step\hrule}\vspace{1mm}

\textbf{Prerequisite:} None.

For each of the three datasets ( NSL-KDD, CICIDS-2017, and RoEduNet-SIMARGL2021), the Efficiency metric is applied using IG, LRP, or DeepLift XAI methods as follows:

\begin{enumerate}
 	\item After training the DNN model, apply IG, LRP, or DeepLift for N samples. We use 1, 100, 500, 2500, and 10000 samples.
  
    \item Track the time required to complete the XAI evaluation method for each run.
        \vspace{1mm}

  \hrule
    \vspace{2mm}
\end{enumerate}

\noindent \textbf{C.4. Stability Step-by-Step\hrule}
\vspace{1mm}
\textbf{Prerequisites:} None

For each of the datasets (CICIDS-2017, NSL-KDD, and RoEduNet-SIMARGL2021), the Stability metric is applied  using IG, DeepLift, and LRP XAI methods at this moment:

\begin{enumerate}

    \item For each dataset, select the top-$k$ features to compare in each run, depending on the total number of features. For example, CICIDS-2017 has over 70 features, so comparing the top $k$ = 20 features is representative. 

    \item Check stability using the formula: $$ Stability = \frac{\bigcap_{i=1}^{N} X_i}{k},$$
    where $N$ is the number of runs, and $k$ is the number of top features. We evaluate how many times the top-$k$ features are common in different $N$ runs. If all the X sets are the same, the stability score is 1; if none are the same, the score is 0.

    \item Repeat the process for $N$ = 3 (three runs or trials) for each AI model and intrusion detection dataset.    \vspace{1mm}
\hrule
    \vspace{2mm}
\end{enumerate}

\noindent \textbf{C.5. Robustness Step-by-Step\hrule}
\vspace{1mm}
\textbf{Prerequisites:} Algorithm from paper~\cite{adversary}.

For each intrusion dataset (RoEduNet-SIMARGL2021, CICIDS-2017, and NSL-KDD), run the Robustness experiment hereby:

\begin{enumerate}
    
	\item Train the biased and adversarial models to decouple the original explanation from the post-hoc XAI explanation (by IG, LRP, or DeepLift).
 
	\item For the adversarial model, use the top-$k$ (k= 6) intrusion features for the unrelated feature. For the Biased model, use only the Biased feature.

	\item Use IG, LRP, or DeepLift to verify if the Biased feature appears as the top feature in the biased model. For the adversarial model, check if the unrelated and biased features appear in the top-3 positions.

    \item Run the Robustness metric for 100 samples and stack the top-3 features in a bar graph for each run and model (adversarial and biased).
    
    \item Assess the occurrence of each feature in the text. The biased model probably will have the biased feature as the top-1 feature.
    
    \item For the adversarial model, check if the biased and unrelated features are in the top 3 features. The presence of the unrelated feature in a higher rank than the biased feature indicates the explanation was compromised.\hrule
\vspace{1mm}
\end{enumerate}

\textbf{Discussion:} The primary point of robustness is demonstrating that an adversarial attack on XAI methods (as explained above) is possible. We do not delve into the details of performing the attack, such as breaking into a system and replacing the correct AI model with an adversarial one to generate the disguised explanation. The attacker exploits the fact that IG, LRP, or DeepLift are post-hoc XAI methods. However, it is worth mentioning their superior robustness given that the model itself is an input for the analysis, making it more resistant to perturbations and easier to troubleshoot.

\noindent \textbf{C.6. Completeness Step-by-Step} \hrule \vspace{1mm} We specify below the step-by-step process for local completeness (on a single instance) and global completeness (on various instances) for the completeness XAI evaluation metric used in this paper.

\noindent \textbf{C.6.1 Local Completeness (Single instance) Step-by-Step:\hrule}
\vspace{1mm}
\textbf{Prerequisites:} None

For each of the datasets (NSL-KDD, CICIDS-2017, and RoEduNet-SIMARGL2021), the Completeness metric is applied using DNN in tandem with IG, LRP, or DeepLift XAI methods as follows:

\begin{enumerate}

    \item Divide the dataset by class type to assess the results by label.

    \item Produce an explanation with IG, LRP, or DeepLift and a prediction for the chosen network traffic instance.

    \item Starting with the top feature in descending order, perturb the value of the feature in the instance from 0 to 1.0 in small steps of 0.1.

    \item For each perturbation, produce a new prediction and explanation with IG, LRP, or DeepLift.

    \item If the new prediction differs from the original one, stop the process. A class change indicates that the explanation was valid.

    \item If the new prediction is the same as the original, repeat the process for the second most important feature (Steps 4 and 5).\hrule
    \vspace{2mm}
\end{enumerate}

\noindent \textbf{C.6.2 Global Completeness Algorithm \hrule} \vspace{1mm} The proposed algorithm for testing global completeness (for various instances) is shown hereby:
\begin{enumerate}
    \item Split the batches by class.
    \item Produce the explanation for the original instance.
    \item Perturb the top-1 feature and check if the class changes. Repeat the process for subsequent features up to the top-k feature. The perturbation changes the feature value from 0 to 1 in steps of 0.1. 
    \item If the current perturbed feature fails to change the prediction class, assign its value as $|original value - 1|$ to maintain relevant perturbation values for previously perturbed features. 
    \item If the class changes, stop the experiment for that sample and register the change.
    \item Count how many samples changed class when perturbed and divide this number by the batch size.

    \item Calculate the percentage of samples with valid explanations:
\[
\resizebox{0.9\columnwidth}{!}{$Completeness (\%) = \frac{\text{No. Samples that Changed Class}}{\text{Total Number of Samples}}.$}
\]
\hrule

\end{enumerate}

\subsection{Top Intrusion Features List and Usage in Evaluating XAI}

A list of the top features for the three datasets is presented since they are used often through this work, see Tables~\ref{tbl:feature_list_sensor}-\ref{tbl:feature_list_NSL-KDD_selected}. These tables describe the essential features for each dataset (i.e., RoEduNet-SIMARGL2021, CICIDS-2017, and NSL-KDD), which are invaluable for the evaluation of the used XAI techniques, see Section~\ref{sec:evaluation}. Moreover, such features, as shown in Table~\ref{tab:XAI_topfeatures}, are significant to the XAI experiments, in particular descriptive accuracy, sparsity, and stability. These metrics use the scores of the top features to evaluate their quality. In addition, indirectly, these features are used for the remaining metrics to measure time efficiency and evaluate its correctness and robustness.

\textbf{Important Note:}  The Tables~{\ref{tbl:feature_list_sensor}}-{\ref{tbl:feature_list_NSL-KDD_selected}} presented in this paper display the description of the crucial features for the three datasets. However, the total number of features in Table~{\ref{tbl:samples_distributions_datasets}} were used in the experiments to grasp the full potential of the intrusion datasets. Moreover, the same table contains information about each dataset, including sample size, number of features, and the number of classes available (i.e., intrusion types). Consequentially, the experiments have this information to better leverage them during the experimentation, and as a product of the efforts we present a top feature list for all datasets considering IG, LRP, and DeepLift in Table~{\ref{tab:XAI_topfeatures}}.

\subsection{Application of Features for White-box XAI} 

The feature's list is generated by their XAI methods (i.e., DeepLift, IG, and LRP) and ranked by scores, see Table~\ref{tab:XAI_topfeatures}. The DNN model is trained considering the CICIDS-2017, NSL-KDD, and RoEduNet-SIMARGL2021 datasets, and then LRP, DeepLift, and IG were applied using all features included. The following parts are step-by-step procedures for producing the white-box XAI methods features.
\vspace{2mm}

\textbf{E.1. Feature Generation Steps for LRP and IG (Global)\hrule}

\begin{enumerate}

    \item Select the number of samples to use for generating the global scope and pass it to the iNNvestigate method with the DNN model.

    \item In the same method, pass the parameter as a string (i.e., IG or LRP).
    
    \item iNNvestigate will produce a matrix with the scores for all samples individually. Simply, sum all the scores for each feature to obtain the global feature importance. \hrule \vspace{2mm}
\end{enumerate}

\textbf{E.2. Feature Generation Steps for DeepLift (Global)\hrule}

\begin{enumerate}

    \item Select the number of samples to use for generating the global scope and pass it to the SHAP DeepExplainer method with the DNN model.

    \item Using the SHAP DeepExplainer technique, produce the SHAP Global Summary Plot and extract the list of top features with their relative scores from the graph's properties. \hrule
\end{enumerate}

\begin{table*}[ht]
\centering
\caption{Description of  main features for RoEduNet-SIMARGL2021 dataset \cite{flow1234}.} 

\resizebox{\linewidth}{!}{
\begin{tabular}{l|l}
\hline
\textbf{RoEduNet-SIMARGL2021 Features}               & \textbf{Explanation}      \\                               \hline                                     FLOW\_DURATION\_MILLISECONDS & Flow duration in milliseconds                         \\
PROTOCOL\_MAP                                                           & IP protocol name (tcp, ipv6, udp, icmp)                                                                   \\
TCP\_FLAGS                                                              & Cumulation of all flow TCP flags \\

TCP\_WIN\_MAX\_IN            & Max TCP Window (src-\textgreater{}dst)       \\

TCP\_WIN\_MAX\_OUT           & Max TCP Window (dst-\textgreater{}src) \\

TCP\_WIN\_MIN\_IN            & Min TCP Window (src-\textgreater{}dst) \\

TCP\_WIN\_MIN\_OUT   & Min TCP Window (dst-\textgreater{}src)       \\

TCP\_WIN\_SCALE\_IN  & TCP Window Scale (src-\textgreater{}dst)   \\

TCP\_WIN\_MSS\_IN  & TCP Max Segment Size (src-\textgreater{}dst) \\

TCP\_WIN\_SCALE\_OUT & TCP Window Scale (dst-\textgreater{}src)   \\
SRC\_TOS                                                                & TOS/DSCP (src-\textgreater{}dst)             \\
DST\_TOS                                                                & TOS/DSCP (dst-\textgreater{}src)  \\

FIRST\_SWITCHED & SysUptime of First Flow Packet\\

LAST\_SWITCHED & SysUptime of Last Flow Packet\\
TOTAL\_FLOWS\_EXP & Total number of exported flows
\end{tabular}
}
\label{tbl:feature_list_sensor}
\vspace{-4mm}
\end{table*}

\begin{table*}[h]
\centering
\caption{Description of the main features for the CICIDS-2017 dataset~\cite{ahlashkari_2021}.}
\resizebox{\linewidth}{!}{
\begin{tabular}{l|l}
\hline
\begin{tabular}[c]{@{}c@{}}\textbf{CICIDS-2017  Features}\end{tabular} & \textbf{Explanation}                                                                                                   \\ \hline
Packet Length Std                                         & Standard deviation  length of a packet                                      \\
Total Length of Bwd Packets                               & Total size of packet in backward direction                          \\
Subflow Bwd Bytes                                         & Average number of bytes in backward sub flow        \\
Destination Port                                          & Destination Port Address                                  \\
Packet Length Variance                                    & Variance length of a packet                               \\
Bwd Packet Length Mean                                    & Mean size of packet in backward direction                           \\
Avg Bwd Segment Size                                      & Average size observed in the backward direction                     \\
Bwd Packet Length Max                                     & Maximum size of packet in backward direction                     \\
Init\_Win\_Bytes\_Backward                                & Total number of bytes in initial backward window \\
Total Length of Fwd Packets                               & Total packets in the forward direction                                                                                \\
Subflow Fwd Bytes                                         & Average number of bytes in a forward sub flow  \\
Init\_Win\_Bytes\_Forward                                 & Total number of bytes in initial forward window \\
Average Packet Size                                       & Average size of packet                                                                                                \\
Packet Length Mean                                        & Mean length of a packet                                                                                               \\
Max Packet Length                                         & Maximum length of a packet                                    
\end{tabular}
}
\label{tbl:feature_list_CICIDs}
\end{table*}

\begin{table*}[t]
\centering
\caption{Description of main features for NSL-KDD~\cite{dhanabal2015study}.}
\resizebox{\linewidth}{!}{
\begin{tabular}{l|l}
\hline
\textbf{NSL-KDD Features} & \textbf{Explanation} \\ \hline
duration & Length of the connection \\
protocol\_type & Type of the protocol (e.g., TCP, UDP) \\
service & Network service on the destination (e.g., HTTP, FTP) \\
flag & Normal or error status of the connection \\
src\_bytes & Number of data bytes from source to destination \\
dst\_bytes & Number of data bytes from destination to source \\
logged\_in & 1 if successfully logged in; 0 otherwise \\
count & Number of connections to the same host as the current connection in the past two seconds \\
srv\_count & Number of connections to the same service as the current connection in the past two seconds \\
dst\_host\_same\_srv\_rate & Rate of connection to the same service on destination host \\
dst\_host\_srv\_count & Count of connections to the same service on destination host \\
dst\_host\_same\_src\_port\_rate & Rate of connections to same source port on destination  \\
dst\_host\_serror\_rate & Rate of connections that have activation flags indicating various types of errors \\
dst\_host\_rerror\_rate & Rate of connections that have rejected flags \\
\end{tabular}
}
\label{tbl:feature_list_NSL-KDD_selected}
\end{table*}

\begin{table*}[ht]
\centering
\caption{Summary and statistics of the three network intrusion datasets used in this work, including the size of the dataset, the number of attack types (labels), and the number of intrusion features. 
}
\resizebox{\linewidth}{!}{
\begin{tabular}{|l|c|c|c|}
\hline
\textbf{Dataset} & \textbf{Number of Labels} & \textbf{Number of Features} &  \textbf{Number of Samples} \\
 \hline
\textbf{CICIDS-2017} & 7 & 78 & 2,775,364 \\ \hline
 \textbf{RoEduNet-SIMARGL2021} 
 & 3 & 29 & 31,433,875  \\ \hline
\textbf{NSL-KDD} & 5 & 41 & 148517 \\ \hline

 \end{tabular}
}
\label{tbl:samples_distributions_datasets}
\vspace{-2mm}
\end{table*}

\section{Foundations of Evaluation}\label{sec:evaluation}

The foundation of our evaluation is presented in this section and it focuses on the questions: (1) Considering the white-box XAI methods, what are the crucial features?; (2) For the proposed metrics, what is the performance of the white-box XAI methods?; (3) What are the strong suits and caveats of the analyzed XAI techniques applied to NIDS?; (4) Which white-box XAI performs the best?; and (5) How does white-box XAI methods compare to black-box XAI? In the next section, the experimental setup is detailed before delving into the results.

\subsection{DataSet Description}

\textbf{NSL-KDD Dataset~\cite{dhanabal2015study}:}  The NSL-KDD is an enhanced version of the KDD dataset~\cite{tavallaee2009analysis} developed as a partnership of the National Research Council of Canada and the University of New Brunswick. It improves on some issues, including a mixture of different intrusion types and normal traffic,  organized from raw traffic data. We use KDDTrain+ for the training part and the KDDTest+ for the testing part~\cite{8rpg-qt98-22}.

\textbf{RoEduNet-SIMARGL2021 Dataset~\cite{mihailescu2021proposition}:} It originated from a partnership of the European Union with the Romanian Education Network through the Horizon program, and it includes real-world (i.e., authentic) data originated from real-time traffic analysis, following a structure that is similar to Netflow~\cite{claise2004cisco}, a CISCO developed protocol for network flow monitoring. 

\textbf{CICIDS-2017 Dataset~\cite{panigrahi2018detailed}:} It embodies different attacks (i.e., Infiltration, Web, Port Scan, DoS, Botnet, Brute Force), created by using a B-profile system~\cite{sharafaldin2018towards}, which is used as simulation tool based on protocols from the network.  This data was created by the Canadian Institute for Cybersecurity at the University of New Brunswick in 2017, serving as a benchmark for intrusion detection.

\textbf{Datasets' Statistics and Summary:} Table~\ref{tbl:samples_distributions_datasets} displays each dataset's size, attack types, and intrusion features.

\vspace{-2mm}

\subsection{Experimental Setup}

\textbf{Computing Resources:} For the performed experiments, we used a high-performance computer (HPC)~\cite{BIGRED200}, supporting demanding machine learning tasks (i.e., performing at approximately 7 petaFLOPs). It has the following characteristics: a single 64-core AMD EPYC 7713 processor (2.0 GHz and 225-watt), 64 GPU-accelerated nodes (each with 256 GB of memory), and four NVIDIA A100 GPUs.

\textbf{Coding Tools:} To leverage open-source tools and maintain an open-source implementation, we chose Python along with various toolboxes (including TensorFlow, ScikitLearn, Pandas, Matplotlib, iNNvestigate, and others).

\textbf{XAI Tools:} We used the following XAI toolboxes:

\textbf{(a) DeepLift (DL)~\cite{DeepLift,DeepLift2,DeepLift3}:} This toolkit utilizes the SHAP DeepExplainer kernel specifically tailored for TensorFlow. According to the SHAP documentation~\cite{SHAP}, this method is an enhanced version of the DeepLift algorithm, allowing us to generate feature explanations for our intrusion detection DNN model. We used this tool to produce local and global explanations.

DeepLift is a popular white-box method that determines the relevance of a prediction by performing a backward pass through the neural network, starting at the output layer and calculating relevance until reaching the input layer. It enforces a conservation law where the relevance assigned to features must sum up to the difference between the input outcome and a reference input. DeepLift supports the explanation of decisions made by feed-forward, convolutional, and recurrent neural networks.

\textbf{(b) Integrated Gradients (IG) ~\cite{IG}:} This tool, part of the iNNvestigate library~\cite{innvestigate}, is another popular white-box XAI method. IG computes explanations using gradients in tandem with a baseline to determine the shortest path between the baseline and the output.

\textbf{(c) Layer-wise Relevance Propagation (LRP)~\cite{LRP}:} Also part of the iNNvestigate library~\cite{innvestigate}, LRP assesses the importance of each input given an output by back-propagating each layer of the neural network according to a set of rules that ensure conservation.

\textbf{Hyper-parameters:} The main hyper-parameter choices for the DNN model are provided in Appendix~\ref{app:ai_models_hyperparams}.

\textbf{Metrics for Evaluating XAI:} The metrics are generated to assess the XAI methods. Figure~\ref{fig:desc_acc_shap} displays each XAI curve for the Descriptive Accuracy experiment. Figure~\ref{fig:sparsity} displays the curves for the sparsity experiment considering the XAI methods. Table~\ref{tab:efficiency} shows the efficiency of each XAI method to generate the explanations ranging from single to many samples. Tables~\ref{globalstability}-\ref{localstability} refer to the stability of the XAI explanations, which were obtained by generating the most important features multiple times, and checking their overlapping.  Figures~\ref{fig:example_robustness}-\ref{fig:IG_ocurrence} refer to the robustness experiment, which was adapted from~\cite{adversary}. It creates an adversary that is used to create an alternative explanation from the noised perturbed original sample to produce a convincing explanation to confuse the security analyst. Tables~\ref{tab:all_completeness_SML}-\ref{tab:all_completeness_NSL-KDD} show the completeness experiment even though the white-box is considered complete by default. By perturbing the most significant features, it is expected to change the predicted class, if it does not change, the explanation is not considered valid. Also, Figure~\ref{fig:samples_remaining} displays how many samples change their prediction after a certain perturbation level. 

\vspace{-2mm}

\subsection{Prelude - Feature Selection}

It is proper to describe how the scores for the significant features were obtained before entering the in-depth evaluation results since they are used throughout the experiments. The scores are obtained considering the different versions of the DNN model in tandem with DeepLift, IG, and LRP global explanations (shown in Appendix~\ref{app:feat}).

Global explanations (for each dataset and DNN model) are needed for the sparsity and descriptive accuracy metrics. These results come in a list form with the feature name and its importance score. For the descriptive accuracy experiment, we removed the top-k features at each iteration. For instance, for the RoEduNet-SIMARGL 2021 dataset, it first removes the top five features and then it checks its accuracy, then the process is repeated until a curve is formed. Note for Sparsity, it is sufficient to have all the features listed with their scores, while for the Descriptive accuracy, the order of features removed is k=0, 10,20,40,80 as the number of features allows.

For Stability, the explanations are generated as well, but it is generated a $p$ number of times, three times for this paper's experiments for stability, and then we evaluated how many of the top features overlap. Since the number of features varies among datasets, we considered the top five for RoEduNet-SIMARGL2021, and the top 20 for NSL-KDD and CICIDS-2017. Considering Efficiency, it matters the most to generate the features and check the computation time for each method (i.e., DeepLift, IG, and LRP) using different sample size sets. 

Robustness is examined under a modified version of~\cite{adversary} to run with DNN, and it is done locally to evaluate the resistance of the XAI technique when facing a perturbation attack many times. In a different setup, Completeness deals with perturbing only the top two features in small increments to check if the explanation can be changed. As an observation, by default, the white-box methods are complete since the model is known, but for the consistency of experimentation and comparison, the experiment is performed as well to verify such claim for white-box XAI methods.


\section{In-Depth Metric Analysis}\label{sec:metric_analysis}

This section shows a detailed metric assessment that involves overall the evaluation results, considering the previous explanation of the experimental setups, metrics, and datasets.

\subsection{Descriptive Accuracy} 
We begin by evaluating the descriptive accuracy of XAI methods on our network intrusion datasets. Recall that descriptive accuracy reflects the significance of intrusion features when they are removed. Specifically, if a feature is crucial to the AI model's decision-making process, the model's prediction accuracy will decline upon the feature's removal. Intuitively, removing more features should result in a greater reduction in prediction accuracy. Graphically, this relationship suggests that removing more features will lead to a smaller area under the curve. Within the context of XAI explanations, a greater drop in prediction accuracy following the removal of a feature indicates a higher explainability power for that feature. Figure~\ref{fig:desc_acc_shap} illustrates the descriptive accuracy for the DeepLift, Integrated Gradients (IG), and Layer-wise Relevance Propagation (LRP) white-box XAI methods across the three network intrusion datasets examined in this study.

\textbf{Axis Explanation:} 
The X-axis values in Figure~{\ref{fig:desc_acc_shap}} represent the number of features removed. A value of zero indicates that no features are removed, meaning all intrusion features are utilized. Conversely, a value of 40 signifies that 40 features are removed, and so on. Features are removed based on their importance as determined by the feature importance lists generated by DeepLift, IG, and LRP XAI methods. Thus, a higher value on the X-axis corresponds to fewer features being used in training and evaluating the model. The Y-axis shows the accuracy of the AI model analyzed under each feature setup.

\textbf{Main Insights:} 
Our experiments demonstrate that LRP and IG outperform DeepLift in terms of global explainability, as illustrated in Figure~\ref{fig:desc_acc_shap}. When analyzing XAI curves for descriptive accuracy, most cases exhibit the expected behavior, with accuracy decreasing as more intrusion features are removed, resulting in a slope in the curves. The ideal scenario entails an exponential decay in accuracy, indicating the importance of the extracted features by XAI methods. However, not all cases conform to this expectation, as some do not show a drop in accuracy when features are removed.
Specifically, our experiments reveal that the RoEduNet-SIMARGL2021 dataset exhibits the highest number of XAI methods with an accuracy drop, with both LRP and IG showing a decrease from 0.7 to 0.3 approximately. Similarly, the NSL-KDD dataset follows a trend of decreasing accuracy, particularly for LRP. However, the CICIDS-2017 dataset does not consistently show a continuous accuracy drop for LRP; instead, only IG and DeepLift achieve the expected results, with the accuracy drop ranging between 0.85 and 0.79.

LRP consistently outperformed IG and DeepLift in terms of descriptive accuracy, as illustrated in Figure~\ref{fig:desc_acc_shap}. However, the LRP curve for the CICIDS-2017 dataset shows unusual behavior, with accuracy remaining the same or even increasing when top features are removed. One possible explanation for this phenomenon is that all intrusion features contribute nearly equally to the DNN model. 
In neural network models, this behavior can be attributed to the "curse of dimensionality," where there are significantly more features than predictions. This can lead to overfitting or an inability to effectively model observations with so many features. In our application's context, the high accuracy despite removing features suggests the presence of latent relationships among the intrusion features, which help maintain accuracy even as features are removed.

\begin{figure*}[t] 
\centering
\begin{subfigure}[t]{.33\textwidth}
\centering
\includegraphics[width=\linewidth]{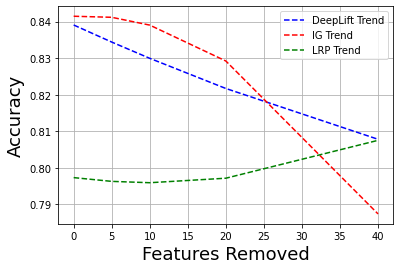}
\vspace{-3mm}
   \caption{CICIDS-2017}
\label{fig:SHAP_ACC_CIC}
\end{subfigure}
\begin{subfigure}[t]{.33\textwidth}
\centering
\includegraphics[width=0.99\linewidth]{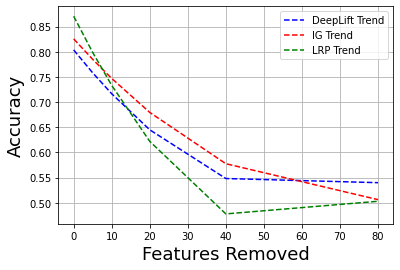}
\vspace{-3mm}
\caption{NSL-KDD}
\label{fig:SHAP_ACC_SML}
\end{subfigure}%
\begin{subfigure}[t]{.33\textwidth}
\centering
\includegraphics[width=0.98\linewidth]{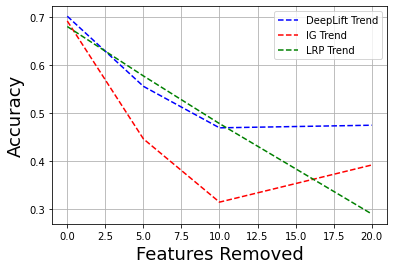} 
   \caption{RoEduNet-SIMARGL}
\label{fig:SHAP_ACC_NSL}
\end{subfigure}
\caption{The Descriptive Accuracy experiment using DeepLift, IG, and LRP white-box XAI methods. The graph displays the accuracy declining as the important intrusion features are removed in the x-axis. It demonstrates
the methods’ effectiveness in global explainability in the three datasets.}
\label{fig:desc_acc_shap}
\end{figure*}

\subsection{Sparsity}
Next, we evaluate the sparsity of white-bpx XAI methods, which reflects the significance of feature importance scores (provided by the XAI method) for different intrusion features. This is done by counting the number of features that fall below a certain threshold, denoted by $\tau$ (ranging from 0 to 1 in increments of 0.1). For example, if at $\tau = 0$, 19 out of 20 intrusion features are below or equal to this threshold, it means only one feature has a significantly high influence on the AI model's decision, indicating high sparsity. Graphically, higher sparsity corresponds to more area under the curve. The idea is that a higher area under the curve (AUC) shows that the XAI method needs fewer top features to provide an adequate explanation.

\textbf{Axis Explanation:}
The X-axis values in Figure~{\ref{fig:sparsity}} represent thresholds that increase in increments of 0.1, ranging from 0 to 1. These thresholds are set manually to compare with the importance scores from DeepLift, LRP, and IG, determining how many features are below the threshold at a given point. Note that higher importance scores indicate greater feature significance according to the XAI technique. For instance, if we consider a total of 10 features and find that 3 out of 10 features have scores less than 0.5, the sparsity score would be 0.3 when the threshold is set to 0.5. Since the importance scores from the XAI methods are normalized using min-max scaling, this metric equals one when the threshold value is one, as all features will have an importance score of at most one. A peak on the left-hand side of the graph represents the ideal scenario, indicating that most features have low explanation potential, concentrating the explanation potential in a small set of intrusion features.

Conversely, the Y-axis represents the sparsity score, which ranges from zero to one. This score is calculated by counting the number of features with importance scores below or equal to each threshold value on the x-axis, and then dividing this count by the total number of features.

\textbf{Main Insights:} 
Figure~\ref{fig:sparsity} illustrates the sparsity for the DeepLift, IG, and LRP white-box XAI methods across all three network intrusion datasets. IG and LRP demonstrate slightly better performance in terms of sparsity compared to DeepLift. Specifically, the curves representing LRP and IG show a higher area under the curve compared to DeepLift for CICIDS-2017, indicating that their sparsity is more pronounced with smaller thresholds. Conversely, DeepLift shows a marginal advantage for the RoEduNet-SIMARGL-2021 dataset compared to IG and LRP, although the overall performance is similar. In contrast, this situation is reversed for NSL-KDD. Overall, Figure~\ref{fig:sparsity} demonstrates that IG and LRP yield the best results for the CICIDS-2017 and NSL-KDD datasets, while DeepLift is more effective for RoEduNet-SIMARGL-2021. The exponential growth shape of the curves indicates that the explanation is concentrated in only a few top intrusion features, suggesting higher explainability potential for these datasets. Given the similar performance for NSL-KDD and RoEduNet-SIMARGL-2021, one could use CICIDS-2017 as a benchmark and leverage IG and LRP as the best performers in this category.

\begin{figure*}[h]
\begin{subfigure}[t]{.33\textwidth}
\centering
\includegraphics[width=\linewidth]{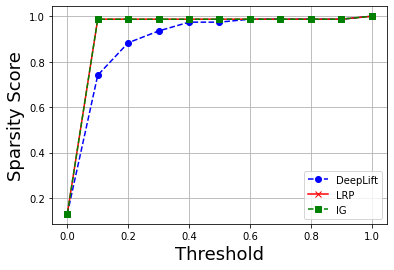}
\vspace{-3mm}
   \caption{CICIDS-2017
}
\label{fig:LIME_ACC_CIC}
\end{subfigure}%
\begin{subfigure}[t]{.33\textwidth}
\centering
\includegraphics[width=0.99\linewidth]{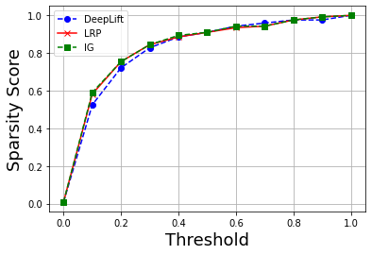}
\vspace{-3mm}
   \caption{NSL-KDD}
\label{fig:LIME_ACC_SML}
\end{subfigure}%
\begin{subfigure}[t]{.33\textwidth}
\centering
\includegraphics[width=0.995\linewidth]{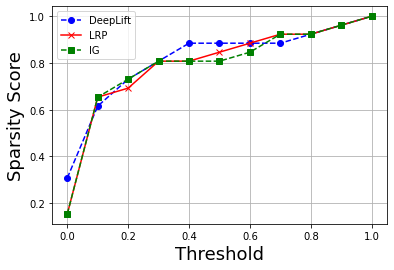}
\vspace{-3mm}
   \caption{RoEduNet-SIMARGL}
\label{fig:LIME_ACC_NSL}
\end{subfigure}
\caption{The XAI techniques Sparsity plots considering LRP, IG, and DeepLift for the used datasets. The outcomes display comparable performance for the datasets. However, in the CICIDS-2017 case, IG and LRP show best performance.}
\label{fig:sparsity}
\end{figure*}

\subsection{Stability} 
Next, we assess the stability of various XAI methods. The stability experiment involved running each XAI explanation three times. Stability is measured by calculating the percentage of features that intersect across the three runs relative to the total number of features. The stability range is from zero to one, with a value closer to one indicating higher stability and reliability (i.e., consistent explanations across identical runs). We evaluate stability both globally (across multiple traffic instances) and locally (on single traffic instance).

\textbf{Main insights:} As anticipated, local stability is higher than global stability because it assesses the same instance, whereas a collection of samples is more likely to have differences by chance. Our experimental results (see Tables~\ref{globalstability}-\ref{localstability}) show that LRP, IG, and DeepLift perform similarly within the local scope. For the global scope, all three XAI methods perform equally well on the NSL-KDD and RoEduNet-SIMARGL2021 datasets. The difference arises when comparing the global results for the CICIDS-2017 dataset: LRP shows the best performance, followed by DeepLift, with IG in the last position. However, the performance difference is not substantial; LRP achieved a stability score of 0.65, while IG scored 0.55. Therefore, LRP is marginally better in this experiment.

\begin{table*}[t!]
\centering
\caption{The global stability evaluation metric of white-box XAI methods. We measure such stability by calculating
the percentage of features that intersect in different runs among the total
number of features for several traffic instances.}
\resizebox{\linewidth}{!}{
\begin{tabular}{l|ccc}
\hline
\textbf{White-box Method}  & \textbf{CICIDS-2017} & \textbf{RoEduNet-SIMARGL2021} & \textbf{NSL-KDD} \\
\hline
\textbf{DeepLift} & 0.60 & 1.00 & 0.65    \\
\textbf{IG} & 0.55 & 1.00 & 0.65 \\
\textbf{LRP} & 0.65 & 1.00 & 0.65 \\

\hline
\end{tabular}
}
\label{globalstability}
\end{table*}

\begin{table*}[t]
\centering
\caption{The local stability evaluation metric of XAI methods. We measure such stability by calculating
the percentage of features that intersect in different runs among the total
a number of features for a single traffic instance.}
\resizebox{\linewidth}{!}{
\begin{tabular}{l|ccccccc}
\hline
\textbf{White-box Method}  & \textbf{CICIDS-2017} & \textbf{RoEduNet-SIMARGL2021} & \textbf{NSL-KDD} \\
\hline
\textbf{DeepLift} & 1.00 & 1.00 & 1.00    \\
\textbf{IG} & 1.00 & 1.00 & 1.00 \\
\textbf{LRP} & 1.00 & 1.00 & 1.00 \\
\hline
\end{tabular}
}
\label{localstability}
\vspace{-4mm}
\end{table*}

\subsection{Efficiency} 
Next, we assess the efficiency of different white-box XAI methods. The efficiency experiment measures the computational time required to run DeepLift, IG, and LRP in various scenarios without the need to extract any features. Table~\ref{tab:efficiency} presents the efficiency results for these XAI methods across the three network intrusion datasets, both globally and locally. Their computational time remains relatively stable with minor fluctuations. In contrast, DeepLift's computational time increases significantly with the sample size, reaching approximately 20 minutes for 10,000 samples in the CICIDS-2017 and NSL-KDD datasets. This makes DeepLift impractical for real-time IDS when dealing with large sample sizes. Locally, Table~\ref{tab:efficiency} shows that all methods perform similarly in terms of efficiency.

Based on these results, we have the following recommendations from a computational efficiency perspective: (i) All XAI methods are well suited for local analysis, (ii) All XAI methods are well suited for global analysis considering 500 samples, and (iii) LRP and IG are best for global analysis with a higher number of samples.

\begin{table*}[t]
\centering
\caption{The efficiency (amount of time in seconds) for generating IG, LRP, and DeepLift explanations for different datasets and different numbers of samples. Note that DeepLift loses performance as the number of samples increases.} 
\resizebox{\linewidth}{!}
{
\begin{tabular}{l|ccc|ccc|ccc}
\hline
 & \multicolumn{3}{c}{CICIDS}& \multicolumn{3}{|c}{SIMARGL}& \multicolumn{3}{|c}{NSL-KDD}\\
\hline
\textbf{Efficiency (Samples x Seconds)} & \textbf{DeepLift}     & \textbf{LRP}    & \textbf{IG} & \textbf{DeepLift}     & \textbf{LRP}    & \textbf{IG}     & \textbf{DeepLift}     & \textbf{LRP}    &\textbf{IG}     \\
\hline
1 (Local)              & 0.3       & 0.8   & 0.2   &2.15   &14.1   &7.4    & 0.1       & 0.4&0.3\\
100             & 1.0       & 0.8   & 0.2   &3.6    &13.6   &7.7    & 0.8       & 0.4&0.2\\
500             & 8.5       & 0.7   & 0.3   &12.9   &13.1   &7.7    & 6.6       & 0.3&0.3\\
2500            & 95.2      & 0.7   & 0.2   &76.7   &11.7   &6.4    & 79.4      & 0.4&0.3\\
10000           & 1323.7    & 0.9   & 0.4   &651.6  &9.2    &5.3    & 1380.1    & 0.3&0.7\\

\hline
\end{tabular}
}
\label{tab:efficiency}
\vspace{-2mm}
\end{table*}
    
\subsection{Robustness}

The method used for Robustness is an adapted version of~\cite{adversary} to work with the network intrusion datasets (i.e., RoEduNet-SIMARGL2021, CICIDS-2017, and NSL-KDD) and to use DNN models in tandem with the analyzed XAI techniques (i.e., IG, LRP, and DeepLift). In summary, it consists of using two different versions of a model (i.e., Adversarial and Biased) to perform an attack simulation based on perturbing the original sample. Then, it would take advantage of the post-hoc nature of the XAI methods analyzed to swap the explanation. It used the significant features of each dataset to train such models. As an instance, a DoS sample from CICIDS-2017 is heavily characterized by the flow duration feature. Hence, such a feature is used to train the highly biased model, see Figure~\ref{fig:biased_feat}. Meanwhile, an unrelated feature with random values is created and utilized to train the adversarial model, see Figure~\ref{fig:feat_adversarial}. This experiment shows that a maleficent agent could, in theory, perform a DoS assault, while displaying to the analyst a fabricated, yet convincing, explanation perturbed by the random values that could imply normal traffic. In the same fashion, the experiment is conducted for the NLS-KDD dataset and RoEduNet-SIMARGL2021. Additionally, we stress that we adapted the original code to use a deep neural network model instead of the standard random forest from the original code to maintain consistency among the experiments in this article (where our code is shared in the aforementioned link in Introduction). Moreover, the XAI methods IG, LRP, and DeepLift were also included, contrasting the original setup of SHAP and LIME in the prior work~\cite{adversary}.

\begin{figure*}[htp] 
\centering
\begin{subfigure}[t]{.495\textwidth}
   \centering
 \includegraphics[width=\linewidth]{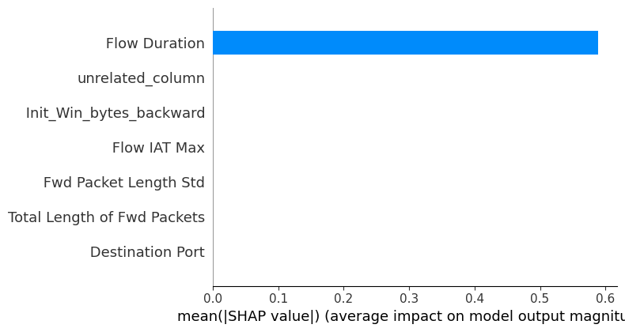}
\vspace{-3mm}
   \caption{Biased}
\label{fig:biased_feat}
\end{subfigure}
\begin{subfigure}[t]{.495\textwidth}
\centering
\includegraphics[width=\linewidth]{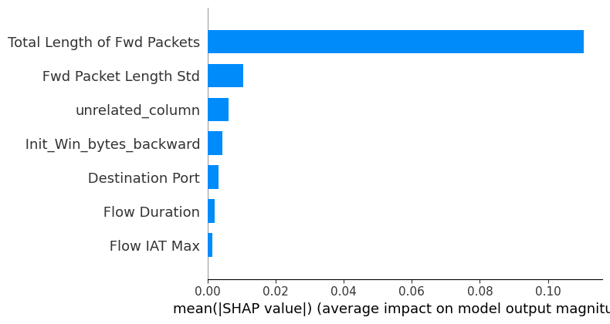}
\vspace{-3mm}
\caption{Adversarial}
\label{fig:feat_adversarial}
\end{subfigure}%
\caption{An illustration of a DoS instance from the CICIDS-2017 dataset, considering the Robustness experiment using DeepLift. In (a), the feature list (with flow duration as the top feature) under a biased explanation is displayed. In (b), the list (with the engineered feature as the top feature) after the adversarial model's classification is exhibited.}
\label{fig:example_robustness}
\end{figure*}

By the white-box XAI methods, IG, LRP, and DeepLift, the experiment is run multiple times (i.e., in this case, 100 times) and organized in the occurrence bar plots. It shows the percentage occurrence of samples for the first three positions (i.e., top three features) for each XAI method. Figure~\ref{fig:DeepLift_ocurrence} refers to the DeepLift Method for all three datasets, and Figures~\ref{fig:IG_ocurrence} and~\ref{fig:LRP_ocurrence} refer to the IG and LRP techniques, respectively. Note that for Figures~\ref{fig:DeepLift_ocurrence}-a,~\ref{fig:IG_ocurrence}-a, and ~\ref{fig:LRP_ocurrence}-a the biased feature is at the first position. Conversely, Figures~\ref{fig:DeepLift_ocurrence}-b, c,d,~\ref{fig:IG_ocurrence}-b, c,d, and ~\ref{fig:LRP_ocurrence}-b, c,d show that that the unrelated feature (i.e., the feature with perturbed values represented as the blue column) appears frequently across the datasets, dividing the top spaces with the biased features that now appear with reduced frequency.

\textbf{Bar Plots:}
\textbf{(1) DeepLift:} 
Figure~\ref{SHAP_Biased} displays the comparison bar plots produced by DeepLift, considering the datasets analyzed. It is noticeable the biased features appear at first.
The best result is for the RoEduNet-SIMARGL2021 dataset (Figure~\ref{SHAP_Adversarial_SML}), where the biased feature almost always appears in the third position, with minimal influence from the unrelated feature (blue column), indicating robustness against random values. The next best result for DeepLift is CICIDS-2017 (Figure~\ref{SHAP_Adversarial_CIC}), where the biased feature appears in the second and third positions about 30\% of the time, though it is somewhat affected by random values from the adversarial model. The NSL-KDD experiment with DeepLift (Figure~\ref{SHAP_Adversarial_NSL}) ranks last due to the sparse appearance of the biased feature but shows potential with low influence from the unrelated column.

\textbf{(2) IG:} Comparing the bar plots from IG for the three datasets, the biased feature consistently takes the top spot in all occurrences (see Figure~\ref{fig:IG_ocurrence}). The best result comes from the CICIDS-2017 dataset (Figure~\ref{IG_Adversarial_cic}), 
where the biased feature occupies the first and second positions around 40\% of the time, despite some susceptibility to the unrelated (blue) column. The RoEduNet-SIMARGL2021 experiment, (Figure~\ref{IG_Adversarial_sml}) shows an intermediate result, with the biased feature in the second position about 60\% of the time but also affected by the unrelated column. The NSL-KDD dataset (Figure~\ref{IG_Adversarial_nsl}) demonstrates higher robustness to the unrelated column but uncovers the biased feature less frequently compared to the other datasets.

\textbf{(3) LRP:} Comparing the bar plots from LRP for the three datasets, the biased feature consistently takes the top spot in all occurrences (see Figure~\ref{fig:LRP_ocurrence}). The CICIDS-2017 dataset (Figure~\ref{LRP_Adversarial_cic}) shows the best results for LRP, with the biased feature appearing in all three positions, most often in the third, followed by the first and second positions, though it is somewhat susceptible to random values as indicated by the frequent appearance of the blue column. This susceptibility to the unrelated column is also evident in the RoEduNet-SIMARGL2021 experiment (Figure~\ref{LRP_Adversarial_sml}) with the biased feature appearing less frequently than in the previous experiment. However, it still performs better than the NSL-KDD dataset (Figure~\ref{LRP_Adversarial_nsl}) where the biased feature appears in any position less than 10\% of the time.

\textbf{Main Insights:} Several conclusions can be drawn from these robustness experiments. Firstly, IG and LRP show the best results across all three datasets, with IG slightly outperforming LRP. However, for RoEduNet-SIMARGL2021, one could argue that DeepLift is superior due to its increased robustness against the unrelated column and nearly 100\% identification of the biased feature in the third position. The results also indicate that methods applied to the NSL-KDD dataset tend to show higher robustness but are less effective at uncovering biased features, with DeepLift being the least affected by the unrelated column due to its low overall occurrence. Therefore, based on its overall performance in most scenarios, IG is considered the most robust white-box XAI method in this experiment.

\begin{figure*}[t] 
\centering
\begin{subfigure}[t]{.24\textwidth}
   \centering
\includegraphics[width=\linewidth]{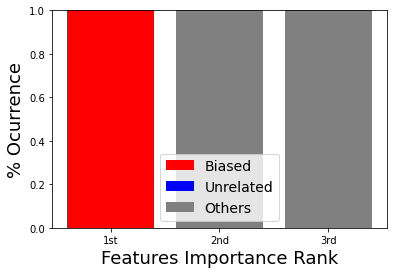}
\vspace{-3mm}
\caption{Biased}
\label{SHAP_Biased}
\end{subfigure}
\begin{subfigure}[t]{.24\textwidth}
\centering
\includegraphics[width=\linewidth]{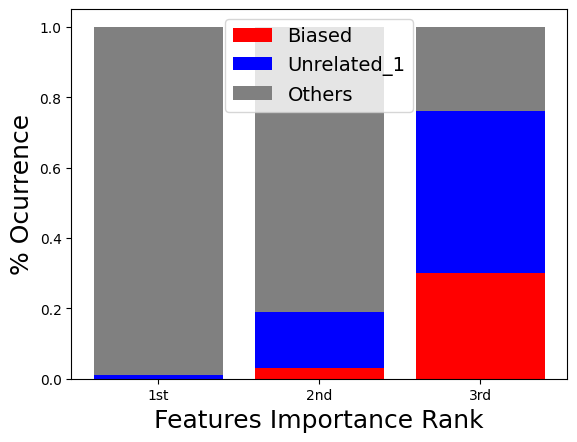}
\vspace{-3mm}
\caption{DeepLift - Advers CIC.}
\label{SHAP_Adversarial_CIC}
\end{subfigure}%
\begin{subfigure}[t]{.24\textwidth}
   \centering
\includegraphics[width=\linewidth]{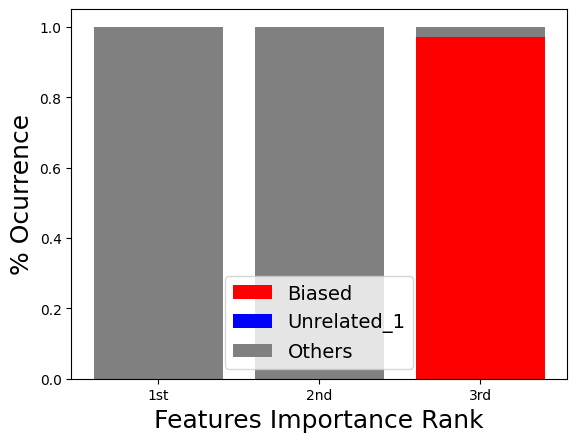}
\vspace{-3mm}
\caption{DeepLift - Advers SML.}
\label{SHAP_Adversarial_SML}
\end{subfigure}
\begin{subfigure}[t]{.24\textwidth}
\centering
\includegraphics[width=\linewidth]{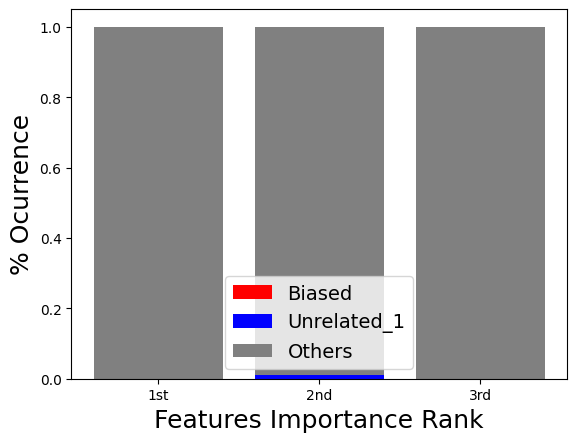}
\vspace{-3mm}
\caption{DeepLift - Advers NSL.}
\label{SHAP_Adversarial_NSL}
\end{subfigure}%

\caption{The percentage of data samples for which biased and unrelated features appear in top-3 features (according to DeepLift rankings of feature importance) for the biased classifier (in (a)) and adversarial classifier (in (b), (c) and (d)) that uses one uncorrelated feature for each dataset. 
Note that (c) displays the best result. It barely suffers the influence of the unrelated column while displaying the Biased Feature in the third position.}
\label{fig:DeepLift_ocurrence}
\vspace{-1mm}
\end{figure*}

\begin{figure*}[t] 
\centering
\begin{subfigure}[t]{.24\textwidth}
   \centering
\includegraphics[width=\linewidth]{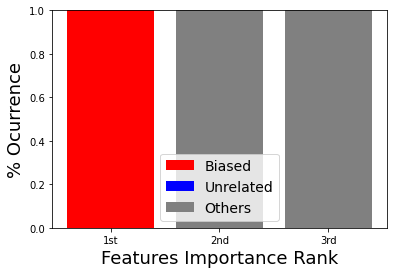}
\vspace{-3mm}
\caption{Biased}
\label{SHAP_Biased_sml}
\end{subfigure}
\begin{subfigure}[t]{.24\textwidth}
\centering
\includegraphics[width=\linewidth]{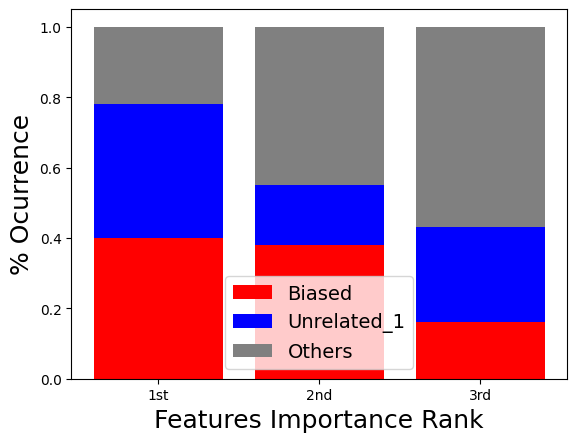}
\vspace{-3mm}
\caption{IG - Advers CIC.}
\label{IG_Adversarial_cic}
\end{subfigure}%
\begin{subfigure}[t]{.24\textwidth}
   \centering
\includegraphics[width=\linewidth]{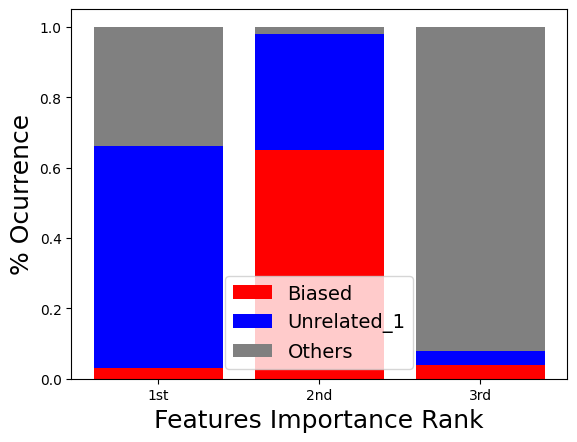}
\vspace{-3mm}
\caption{IG - Advers SML.}
\label{IG_Adversarial_sml}
\end{subfigure}
\begin{subfigure}[t]{.24\textwidth}
\centering
\includegraphics[width=\linewidth]{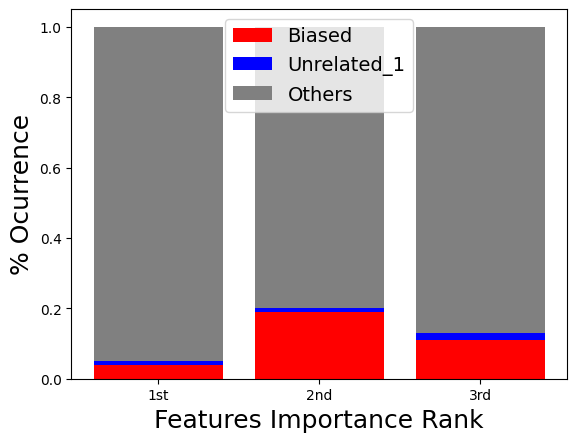}
\vspace{-3mm}
\caption{IG - Advers NSL.}
\label{IG_Adversarial_nsl}
\end{subfigure}%
\caption{The percentage of data samples for which biased and unrelated features appear in top-3 features (according to IG rankings of feature importance) for the biased classifier (in (a)) and adversarial classifier (in (b), (c) and (d)) that uses one uncorrelated feature for each dataset. Note that (b) displays the best result. Although it suffers a relatively high influence from the unrelated column, it displays the Biased Feature in the top three positions most of the time.}
\label{fig:IG_ocurrence}
\vspace{-1mm}
\end{figure*}

\begin{figure*}[t] 
\centering
\begin{subfigure}[t]{.24\textwidth}
   \centering
\includegraphics[width=\linewidth]{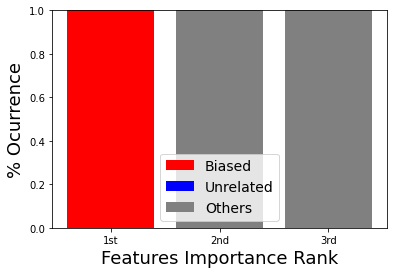}
\vspace{-3mm}
\caption{Biased}
\label{SHAP_Biased_sml_nslkdd}
\end{subfigure}
\begin{subfigure}[t]{.24\textwidth}
\centering
\includegraphics[width=\linewidth]{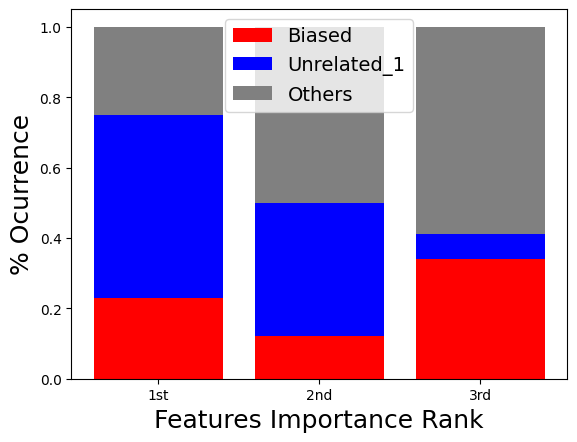}
\vspace{-3mm}
\caption{LRP - Advers. CIC}
\label{LRP_Adversarial_cic}
\end{subfigure}%
\begin{subfigure}[t]{.24\textwidth} 
   \centering
\includegraphics[width=\linewidth]{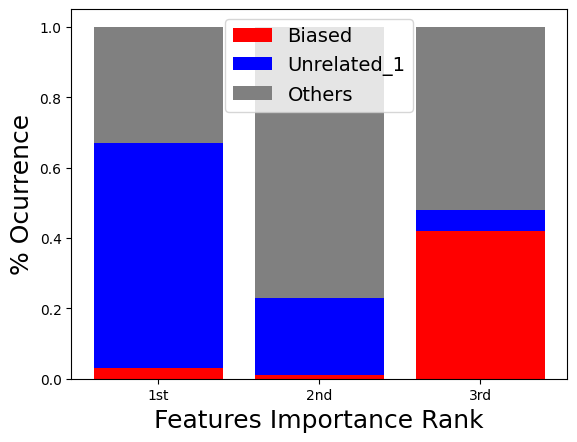}
\vspace{-3mm}
\caption{LRP - Advers. SML}
\label{LRP_Adversarial_sml}
\end{subfigure}
\begin{subfigure}[t]{.24\textwidth}
\centering
\includegraphics[width=\linewidth]{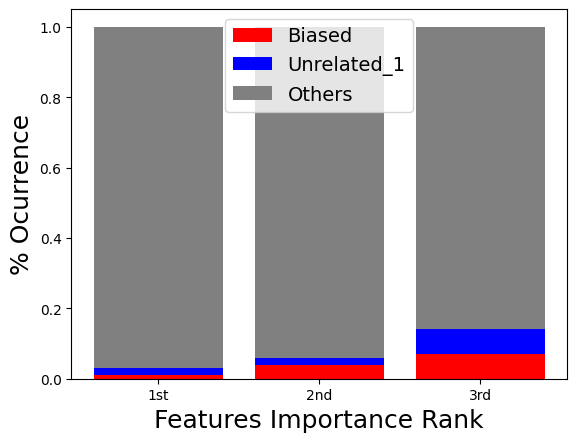}
\vspace{-3mm}
\caption{LRP - Advers. NSL}
\label{LRP_Adversarial_nsl}
\end{subfigure}%
\caption{The percentage of data samples for which biased and unrelated features appear in top-3 features (according to LRP rankings of feature importance) for the biased classifier (in (a)) and adversarial classifier (in (b), (c) and (d)) that uses one uncorrelated feature for each dataset. Note that (b) displays the best result. Although it suffers a relatively high influence from the unrelated column, it displays the Biased Feature in the top three positions most of the time.} 
\label{fig:LRP_ocurrence}
\end{figure*}

\subsection{Completeness}

According to~\cite{evaluating6metrics}, the white-box XAI methods are complete because the parameters and architecture of the model are known and the XAI does not need to rely on approximations. For black-box XAI methods, the model is considered inaccessible. Therefore, the experiment to test its completeness is testing every sample, including corner cases. For the consistency of experimentation, a reduced form is performed on the white-box XAI techniques DeepLift, IG, and LRP. The experiment consists of generating an explanation for the samples and checking its explanation. Then, the top-k features are perturbed with values (e.g., values inside the range of 0 and 1). Lastly, the DNN model classification is rechecked considering the perturbed sample.

The goal is to see if the perturbations can induce a change in the predicted class. If this is the case, it would mean, by this criteria, that the original explanation was relevant and valid. Therefore, if all samples repeat this behavior, it means that the XAI method in question is Complete in the Completeness criteria. Nonetheless, it would take a single non-valid explanation sample to disqualify the XAI method. We performed such analysis on a batch of 100 samples from each class to test completeness for each class and gather meaningful insight into it.

\begin{figure*}
\centering
\begin{subfigure}[t]{.32\textwidth}
   \centering
\includegraphics[width=\linewidth]{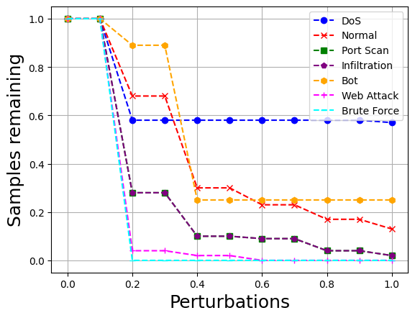}
\vspace{-4mm}
   \caption{DL - CICIDS-2017}
\end{subfigure}
\begin{subfigure}[t]{.32\textwidth}
\centering
\includegraphics[width=\linewidth]{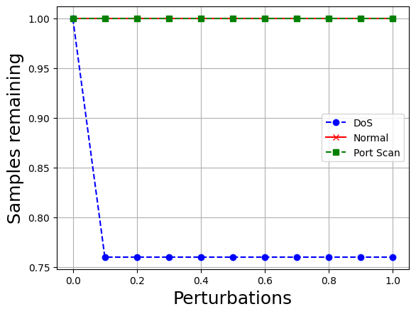}
\vspace{-4mm}
\caption{DL - RoEduNet-SIMARGL2021}
\end{subfigure}
\begin{subfigure}[t]{.32\textwidth}
\centering
\includegraphics[width=\linewidth]{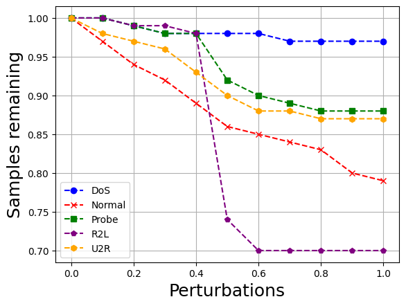}
\caption{DL - NSL-KDD}
\end{subfigure}
\begin{subfigure}[t]{.32\textwidth}
\centering
\includegraphics[width=\linewidth]{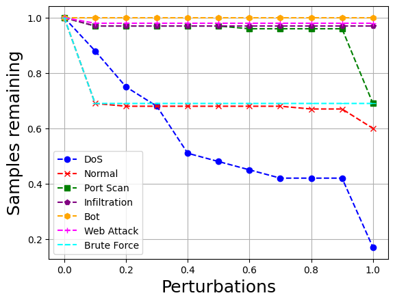}
\vspace{-3mm}
\caption{IG - CICIDS-2017}
\end{subfigure}
\vspace{1mm}
\begin{subfigure}[t]{.32\textwidth}
\centering
\includegraphics[width=\linewidth]{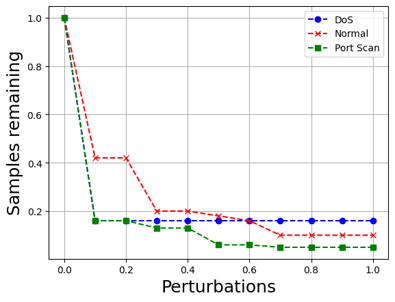}
\vspace{-3mm}
\caption{IG - RoEduNet-SIMARGL2021}
\end{subfigure}
\begin{subfigure}[t]{.32\textwidth}
\centering
\includegraphics[width=\linewidth]{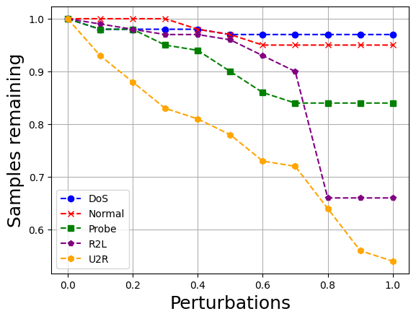}
\caption{IG - NSL-KDD}
\end{subfigure}%

\begin{subfigure}[t]{.32\textwidth}
\centering
\includegraphics[width=\linewidth]{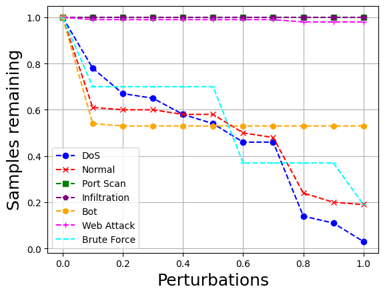}
\vspace{-3mm}
\caption{LRP - CICIDS-2017}
\end{subfigure}
\vspace{1mm}
\begin{subfigure}[t]{.32\textwidth}
\centering
\includegraphics[width=\linewidth]{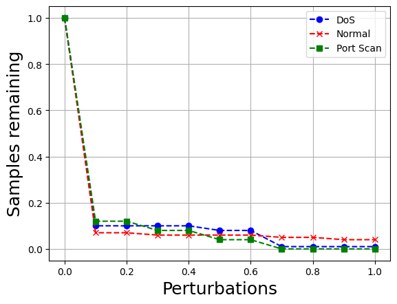}
\vspace{-3mm}
\caption{LRP - RoEduNet-SIMARGL2021}
\end{subfigure}
\begin{subfigure}[t]{.32\textwidth}
\centering
\includegraphics[width=\linewidth]{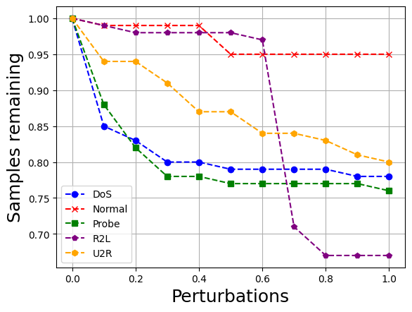}
\caption{LRP - NSL-KDD}
\end{subfigure}%

\caption{Percentage of remaining network traffic samples after removing the samples in which the intrusion class changed under different perturbation levels in intrusion features. We observe that higher perturbations tend to change classes for both DeepLift, IG, and LRP for all three datasets.} 
\label{fig:samples_remaining}
\vspace{-1mm}
\end{figure*}

The setup, as aforementioned, consists of small sets from each attack from each dataset of 100 samples. Each of these samples receives the small increasingly small perturbations in their top features until a class swap is induced. The outcome of this is in Figure~\ref{fig:samples_remaining}. It displays the perturbation effort required to produce a prediction change. Moreover, the y-axis refers to the unchanged samples or the samples that are yet to be changed. Meanwhile, the x-axis refers to perturbation applied to the sample, it increases toward the left. As the perturbation increases, the probability of a class change also increases. In this experiment, DeepLift, and LRP present the most efficient overall behavior, with DeepLift being the best for CICIDS-2017 and RoEduNet-SIMARGL2021, and LRP being the best for NSL-KDD and RoEduNet-SIMARGL2021. Figure~\ref{fig:samples_remaining} displays the attack labels in RoEduNet-SIMARGL2021 have explanations with higher quality than NSL-KDD and CICIDS-2017 because it required a less amount of perturbation to induce the class swap. Analyzing the different scenarios in which the curves did not reach zero (i.e., a complete explanation for that intrusion class), it is due to the insufficient amount of perturbation performed to induce the change, see Tables~\ref{tab:all_completeness_SML}-\ref{tab:all_completeness_NSL-KDD}. For clarification, the results are consistent for three reasons. The first reason is the reduction for this experiment, only the first two features are perturbed, which is not enough to produce a change in several cases. It is stressed this reduced form is due to the demanding memory and time complexity observed empirically when the experiments were conducted. The second reason is the equal or superior performance when compared to the black-box XAI methods~\cite{osvaldo}, which used a more comprehensive form of the experiment. The third reason is the theory that supports white-box XAI~\cite{evaluating6metrics}, in other words, if the explanation is not valid, is highly likely that the AI model needs to be rechecked. 

Moreover, the number of maximum features to be perturbed was chosen to be the top 2 features. The underlying reason is a memory limitation when performing the experiments within LRP and IG which are part of the iNNvestigate package~\cite{innvestigate}. Both methods demand high memory utilization when called multiple times in a short period in a sequential form. Under these conditions, the XAI behavior is to increase drastically the memory consumption and the time to generate the explanations repeatedly. It often led to a kernel crash. Hence, one can assume the iNNvestigate library is not designed to perform under the stress of a completeness experiment.

Tables~\ref{tab:all_completeness_SML}-\ref{tab:all_completeness_NSL-KDD} display the Completeness of the considered datasets for the white-box XAI methods considered in current work (DeepLift, LRP, and IG). As a stressing note, this is the reduced form of the experiment performed for Black-box XAI techniques in~\cite{evaluating6metrics,osvaldo}. Therefore, its principal intuition is to demonstrate the superiority of white-box XAI by enabling a similar ground for comparison, and for this metric backing its theory of the inherent Completeness due to the known model parameters and architectures. As expected, the results display  LRP, IG, and DeepLift XAI methods would not be complete due to the existence of non-valid explanations for the considered intrusion labels because it only uses the top two features due to the memory issues explained in Section~\ref{sec: Discussion}. However, this experiment helps to gather insights. For example, its performance shows greater performance than its black-box XAI counterparts, which supports its theory of white-box XAI methods being inherently complete.

In other words, in this reduced form of the experiment, DeepLift, LRP, and IG failed to produce a change in class after the applied perturbations in the top two features. Although the predicted class did not change, the explanations changed with the perturbation, meaning that if the original top two features were significant, it would have produced a higher impact on the model prediction. 
For the RoEduNet-SIMARGL2021 dataset, Table~\ref{tab:all_completeness_SML} displays its incompleteness for Port Scan and Normal instances, given DeepLift. However, IG and LRP display a high completion (i.e., over 84\% complete for all classes). Plus, Table~\ref{tab:all_completeness_SML} displays LRP having complete cases for the whole 100 Port Scan instances in the experiment. Table~\ref{tab:all_completeness_CIC} shows the outcomes for the CICIDS-2017 dataset. Contrary to Table~\ref{tab:all_completeness_SML}, which shows LRP and IG performed best, Table~\ref{tab:all_completeness_CIC} displays DeepLift having the highest performance on the CICIDS-2017 dataset (i.e., Complete explanations for higher assault labels). Considering, that the same technique was used for perturbing the features, it can be concluded that LRP and IG tend to lower their performance in terms of completeness, considering the scenario of multi-classification (i.e., more intrusion attacks, which is the case with CICIDS-2017 dataset).

For the NSL-KDD dataset, Table~\ref{tab:all_completeness_NSL-KDD} shows that the performance of DeepLift, IG, and LRP differs significantly for this dataset. For the normal class, DeepLift performs noticeably better than LRP and IG. Meanwhile, LRP is the best performer for the DoS and Probe classes, and IG performs best when dealing with U2R and R2L. This discrepancy confirms that in this scenario each XAI method is best for a different class. To sum up, our experiments highlight that LRP and DeepLift are more complete than IG, indicating that they can provide more thorough and dependable interpretability in intricate classification scenarios. In addition, DeepLift has the advantage of not having the memory constraints for this experiment.

\begin{table*}[t!]
    \centering
        \caption{The percentage of samples that are complete (i.e., with valid explanations) for each intrusion class used for the RoEduNet-SIMARGL2021 dataset. LRP has the best results.
    }
    \resizebox{0.5\linewidth}{!}{
    \begin{tabular}{c|ccc}
    \hline
         & \textbf{Normal} & \textbf{DoS} & \textbf{Port Scan}\\
         \hline
         \textbf{DeepLift} & 0\% & 24\% & 0\%\\
         \textbf{IG} &  90\%&  84\%& 95\%\\
         \textbf{LRP} &  96\%&  99\%& 100\%\\
    \end{tabular}
    }
\label{tab:all_completeness_SML}
\end{table*}

\begin{table*}[t]
    \centering
       \caption{The percentage of samples that are complete (i.e., with valid explanations) for each intrusion class using the CICIDS-2017 dataset. DeepLift obtained the best results}
    \resizebox{0.7\linewidth}{!}
{
    \begin{tabular}{c|cccc}
    \hline
         & \textbf{Normal} & \textbf{DoS} & \textbf{Port Scan} & \textbf{Bot}   \\
         \hline
         \textbf{DeepLift}  &   87\%    &   43\%    &98\% & 75\%  \\
         \textbf{IG}        &   40\%    &   83\%    & 31\% & 0\%\\
         \textbf{LRP}       &   81\%    &   97\%    & 0\% & 47\%\\
         \hline
          & \textbf{Brute Force} & \textbf{Infiltration} & \textbf{Web Attack}  &    \\
          \hline
          \textbf{DeepLift}     & 100\% &  98\% & 100\% &    \\
         \textbf{IG}            &  31\% &  3\% & 2\% &   \\
         \textbf{LRP}           &  81\% &  0\% & 2\% &   \\
          
    \end{tabular}
\label{tab:all_completeness_CIC}
}
\end{table*}

\begin{table*}
    \centering
    \caption{The percentage of samples that are complete for each class using the NSL-KDD dataset. Results are mixed and class-dependent.}
\label{tab:all_completeness_NSL-KDD}
\resizebox{0.7\linewidth}{!}{
    \begin{tabular}{c|ccccc}
    \hline
         & \textbf{Normal} & \textbf{DoS} & \textbf{Probe} & \textbf{U2R} & \textbf{R2L} \\
         \hline
         \textbf{DeepLift}  & 21\% & 3\% & 12\% & 13\% & 30\%\\
         \textbf{IG}        & 5\% & 3\% & 16\% & 46\% & 34\%\\
         \textbf{LRP}       & 5\% & 22\% & 24\% & 20\% & 33\%\\
    \end{tabular}
    }
\end{table*}

\begin{table*}
    \centering
     \caption{A summary of performance comparison between Deeplift, IG, and LRP for all six performance metrics. Overall, SHAP provides the best performance compared to LIME for the six metrics analyzed in this work.}
\resizebox{\linewidth}{!}{
    \begin{tabular}{c|c|c|c|c|c|c}
    \hline
       \textbf{CICIDS-2017}& \textbf{Descriptive} \textbf{Accuracy} & \textbf{Sparsity} & \textbf{Stability} & \textbf{Completeness} & \textbf{Robustness} & \textbf{Efficiency} \\
       \hline
       \textbf{DeepLift}& & & & \checkmark & & \\
        \textbf{IG}&  \checkmark &  \checkmark & &  & \checkmark & \checkmark\\
 \textbf{LRP}& \checkmark & \checkmark & \checkmark & & &\\
        \hline

 \textbf{RoEduNet-SIMARGL2021}& \textbf{Descriptive} \textbf{Accuracy} & \textbf{Sparsity} & \textbf{Stability} & \textbf{Completeness} & \textbf{Robustness} &
\textbf{Efficiency} 
\\
\hline
 \textbf{DeepLift}& & \checkmark 
& \checkmark & \checkmark & \checkmark &

\\
 \textbf{IG}& \checkmark & & \checkmark & & \checkmark &\checkmark \\
 \textbf{LRP}& \checkmark & & \checkmark & \checkmark & &\\
        \hline

 \textbf{NSL-KDD}& \textbf{Descriptive} \textbf{Accuracy} & \textbf{Sparsity} & \textbf{Stability} & \textbf{Completeness} & \textbf{Robustness} &\textbf{Efficiency} 
\\
\hline
 
\textbf{DeepLift}& & & \checkmark & & &\\
 \textbf{IG}& & \checkmark 
& \checkmark & & \checkmark &
\\
 \textbf{LRP}& \checkmark & \checkmark & \checkmark & \checkmark & &

\checkmark \\

        \hline
        \hline
 \textbf{Metric - Overall}& \textbf{Descriptive} \textbf{Accuracy} & \textbf{Sparsity} & \textbf{Stability} & \textbf{Completeness} & \textbf{Robustness} &\textbf{Efficiency} 
\\
\hline
 
\textbf{DeepLift}
& & & & \checkmark & &\\
 \textbf{IG}
& & \checkmark & & & \checkmark &
\checkmark \\
 \textbf{LRP}& \checkmark & \checkmark & \checkmark & \checkmark & &\\

    \end{tabular}
   }
    \label{tab: Summary Metrics Table}
\end{table*}

\textbf{Summary of Results:}  Table~\ref{tab: Summary Metrics Table} summarizes the results for both white-box XAI techniques (IG, LRP, and DeepLift) concerning the six XAI evaluation metrics considered in this work. Overall, LRP achieved the best results being the best performer in Descriptive Accuracy, Sparsity, Stability, and Completeness. IG performed best for Sparsity, Robustness, and Efficiency. Lastly, DeepLift performed better for Completeness. Nonetheless, Table~\ref{tab: Summary Metrics Table}, also shows the performance per dataset. For instance, IG is the best performer for the CICIDS-2017 dataset, followed by LRP and DeepLift. Such results could indicate a tendency towards LRP being superior, but also that the individual characteristics of the datasets cannot be ignored.

\subsection{Overall Discussion of Results} 

The evaluation results are favorable to the use of white-box XAI frameworks with the potential of being a step in the direction of applying such techniques in real-world IDS scenarios. The statement is based on the evaluation of the six proposed metrics, which cover the security and AI aspects, and the application of three invaluable intrusion detection datasets in tandem with DNN models. These datasets are comprehensive, RoEduNet-SIMARGL2021 is considered a real-world dataset with relevant traffic data, while CICIDS-2017 and NSL-KDD are highly-regarded benchmark datasets. In addition, LRP, IG, and DeepLift are powerful white-box XAI methods that performed highly in this study. Hence, the comprehensive scenario displays the potential of this work to push forward the use of white-box XAI in real-world scenarios. Lastly, all codes used are open-sourced to the community.

Nonetheless, the outcomes of this research note that in the current form, it is not advised to use DeepLift, LRP, or IG in a production environment yet. The rationale is the need for improvement in some metrics for the XAI methods. For instance, it is not desirable for the security analyst to wait for an explanation while an assault might be happening. Conversely, if the XAI is easily disrupted by perturbations, either because of an attack or from the network, it could impair the thrust in the system. More on the Robustness experiment, the outcomes show that a maleficent agent could in theory take advantage of the post-hoc XAI explanation technique, intercepting or switching the generated explanation. It is crucial to point out this risk, and that the security community is alert to the issue, possibly trying to remedy it through different techniques, or even enhancing DeepLift, IG, and LRP. In the current state and implementation, it is displayed in this study, that such techniques could be exploited in these scenarios.

\subsection{White-box XAI Methods results vs Black-box XAI Methods Results}

In this subsection, we discuss the results differences between this work and our previous paper~\cite{osvaldo}. First, this work uses a DNN model, while ~\cite{osvaldo} uses DNN, LGBM, ADA, RF, KNN, SVM, and MLP. Also, the previous work~\cite{osvaldo} uses SHAP and LIME (Black-box XAI), while this one uses LRP, IG, and DeepLift (White-box XAI). The current XAI methods used are model-dependent and only work with models based on TensorFlow, and other neural network approaches, such as PyTorch. Moreover, Section~\ref{sec:evaluation} details the differences between the mentioned approaches in depth. However, the databases and the criteria used are the same, which allows the comparison between works and how each XAI method performs in each one of the metrics.

\textbf{(1) Descriptive Accuracy:}
   (i) NSL-KDD: For this dataset, all white-box XAI methods outperformed LIME (with all AI models), and SHAP (but RF and LGBM).

   (ii) CICIDS-2017: For this dataset, SHAP paired with SVM is the best performer among all the other XAI methods, followed by IG, SHAP paired with RF, and DeepLift. 

    (iii) RoEduNet-SIMARGL2021: For this dataset, the best performers are IG, DNN, and MLP with SHAP, followed by LRP, DeepLift, and DNN with LIME.

\textbf{(2) Sparsity:}
    (i) NSL-KDD: For this dataset, all white-box XAI methods outperformed LIME (with all AI models), and SHAP (but LGBM).

    (ii) CICIDS-2017:  For this dataset, all LRP and IG outperformed LIME and SHAP (with all AI models), and DeepLift has similar performance with some of SHAP methods, such as the one paired with SVM or RF. 

    (iii) RoEduNet-SIMARGL2021:  For this dataset, all white-box XAI methods outperformed LIME (with all AI models), but it did not outperform SHAP in tandem with MLP, LGBM,  DNN, and SVM.

\textbf{(3) Stability:}
    Considering local stability, all white-box XAI methods were as good as or outperformed the black-box XAI methods. Whereas global stability offered mixed results. DeepLift with CICIDS-2017 is better than SHAP-RF, SHAP-DNN, and SHAP-MLP, but it underperformed compared to the rest. LRP with CICIDS-2017 is better or equally better than SHAP-RF, SHAP-DNN, SHAP-MLP, and SHAP-LGBM but it underperformed compared to the rest. For the NSL-KDD dataset, all the white-box XAI methods underperformed when compared to SHAP and LIME. Lastly, DeepLift, IG, and LRP with RoEduNetSIMARGL2021 performed equally or better than its black-box XAI methods counterpart.

\textbf{Efficiency:}
    For local efficiency, all white-box XAI methods perform worse than their SHAP counterparts, but they outperform all of their LIME counterparts. 

    For Global efficiency, all white-box XAI methods take less than an hour to complete. Hence, it is more efficient compared to LIME. Differently, the LRP and IG (white-box XAI) methods are more efficient than all Black-box XAI methods. DeepLift is more efficient than MLP, SVM, ADA, and KNN paired with SHAP, but equally or inferior to DNN, RF, and LGBM when paired with SHAP. Also, it is more efficient than all LIME experiments.
    
\textbf{Robustness:}
    Overall, all the white-box XAI methods are more robust than the black-box XAI methods. This conclusion can be reached by analyzing the occurrence graphs. Overall, the results show a higher occurrence of the biased feature, and less presence of the unrelated column, which indicates less sensitivity to the random values inserted by such a column. The only exception is LIME paired with NSL-KDD which shows similar performance. 

\textbf{Completeness:}
    Arguably IG and LRP are the best performers for the RoEduNet-SIMARGL2021 dataset because of their high scores for all three classes, while it is slightly inferior compared to SHAP and LIME for DoS and Post Scan, it is fairly superior for the Normal class. Meanwhile, DeepLift shows the worst performance for this dataset. For the CICIDS-2017 dataset, the results are mixed and its class-dependent SHAP shows superior or equal performance for Normal, Brute Force, and Bot, DeepLift is better or equal for PortScan, Web Attack, Brute Force, and Infiltration. LRP is better for DoS. IG has good performance but it does not have the best result for any class, and LIME has the worst performance overall for this metric. For the NSL-KDD dataset, SHAP has the best or equal results for all classes, followed by LRP, IG, DeepLift, and LIME respectively. 

Overall, Table~\ref{tab:black box vs white box} shows the discussion summarized per dataset which concludes that white-box XAI methods are slightly superior in five metrics while staying behind only in the stability performance. 

Having provided the detailed evaluation results, we next discuss the main limitations and prospective future related research directions.

\begin{table*}
    \centering
     \caption{A summary of performance comparison between the black-box and white-box XAI methods for all six performance metrics. Overall, the white-box XAI is slightly superior.}
\resizebox{\linewidth}{!}{
    \begin{tabular}{c|c|c|c|c|c|c}
    \hline
       \textbf{CICIDS-2017}& \textbf{Descriptive} \textbf{Accuracy} & \textbf{Sparsity} & \textbf{Stability} & \textbf{Completeness} & \textbf{Robustness} & \textbf{Efficiency} \\
       \hline
       \textbf{White-box XAI}&               & \checkmark    &                   &  \checkmark           &    \checkmark       & \checkmark \\
        \textbf{Black-box XAI}& \checkmark   &               &  \checkmark       &      \checkmark         &                     & \checkmark\\

        \hline

 \textbf{RoEduNet-SIMARGL2021}& \textbf{Descriptive} \textbf{Accuracy} & \textbf{Sparsity} & \textbf{Stability} & \textbf{Completeness} & \textbf{Robustness} &
\textbf{Efficiency} 
\\
\hline
 \textbf{White-box XAI}&\checkmark   &               &   \checkmark      & \checkmark        &  \checkmark     & \checkmark  \\
 \textbf{Black-box XAI}& \checkmark  & \checkmark    &                   &                   &                 &  \checkmark \\

        \hline

 \textbf{NSL-KDD}& \textbf{Descriptive} \textbf{Accuracy} & \textbf{Sparsity} & \textbf{Stability} & \textbf{Completeness} & \textbf{Robustness} &\textbf{Efficiency} 
\\
\hline
 
\textbf{White-box XAI}   & \checkmark    & \checkmark    &                   &                           &  \checkmark     &  \checkmark\\
 \textbf{Black-box XAI}  &               &               & \checkmark        &  \checkmark       &   \checkmark    &  \checkmark \\

        \hline
        \hline
 \textbf{Metric - Overall}& \textbf{Descriptive} \textbf{Accuracy} & \textbf{Sparsity} & \textbf{Stability} & \textbf{Completeness} & \textbf{Robustness} &\textbf{Efficiency} 
\\
\hline
 
\textbf{White-box XAI}
&\checkmark & \checkmark&           &\checkmark & \checkmark & \checkmark\\
 \textbf{Black-box XAI}
&\checkmark &     &\checkmark &\checkmark &            & \checkmark
 \\

    \end{tabular}
   }
    \label{tab:black box vs white box}
    \vspace{-3mm}
\end{table*}

\section{Limitations, Discussion, and Future Directions}\label{sec: Discussion}

\textbf{Limitations of iNNvestigate Package:} Some important observations are crucial to highlight regarding the experiments. The iNNvestigate package, the one used to generate the white-box XAI methods IG and LRP, cannot work with TensorFlow models when it uses softmax. The lack of the softmax layer can impact the final accuracy of the DNN model, which probably impacts some of the metrics, such as Descriptive Accuracy or Completeness, that could have been better if the softmax layer had been used.  Another limitation is that IG and LRP are memory intensive under stress conditions performed for the Completeness experiments. It was perceived that repeatedly invoking LRP and IG for different perturbation scenarios made the program increase in memory and time complexity, leading to many hours to experiment with and often crashes by the memory demand. This led to the use of less samples and fewer features perturbed this experiment in particular and undermined its performance for this metric in particular. Even underperforming, the white-box XAI methods were able to outperform or have similar performance when compared to other black-box XAI models.

\textbf{White-box Completeness:}
Although a completeness experiment was performed for the white-box XAI methods to assert it is completeness, it was to the effect of comparison and consistency with~\cite{osvaldo}. Moreover, the work in~\cite{evaluating6metrics} states that white-box XAI methods are already complete due to their nature and definition because they calculate the importance scores for the relevant features/neurons directly from the weights of the neural network. Also, due to the intrinsic knowledge of the model, there is a complete and valid explanation for every sample. Conversely, black-box XAI methods have an inherent drawback, as a consequence of their lack of knowledge of the model structure which is the unclearness if an incorrect feature explanation occurs because there is a flaw in the model or an issue with the approximations method used to understand the model behavior. Within this context, comparing the results of this piece with the results~\cite{osvaldo} (shown in Table~\ref{tab:black box vs white box}) shows that overall white-box methods are superior for the Completeness metric, even when perturbing only the top two features in comparison to the top 5 features of the black-box methods. As a side explanation, the setup difference is due to the extensive memory usage that the iNNvestigate library (i.e., IG and LRP white-box methods) requires when experimenting.

\textbf{White-box vs. Black-box XAI :} As mentioned before, white-box XAI methods are different from their black-box counterparts. First, white-box methods know the structure and architecture of the model's insides and use it for generating its explanations, whereas black-box XAI methods lack the knowledge of the internal structure of the model, relying on the input/output relationship. Consequentially, this design difference is in tandem with the limitation of white-box methods being model-specific, while black-box methods tend to be model-agnostic. Nonetheless, the results show that the white-box XAI methods are more resilient to attacks when compared to the black-box ones. It under-covered the biased feature more often, and it showed more resistance to perturbation.
 Hence, making these methods a better choice. 

\textbf{The Significance of our XAI Framework for IDS:} Network assaults would naturally increase in frequency in the environment of exponential information expansion that we live in (e.g., see the recent report by the Center for Strategic \& International Studies (CSIS)~\cite{cyber-attacks-decade}). Security analysts are in charge of double-checking potential attack occurrences that are created more quickly than they can be examined, even though IDS has changed over time.  Thus, having precise intrusion detection systems based on XAI can assist in resolving this issue. In this work, we provide a methodology that facilitates a complete evaluation of white-box XAI-based systems before deploying them for network security.

\textbf{Examining our XAI Assessment System on Alternative Benchmark Datasets:} This paper aimed to provide an exhaustive examination of our XAI assessment system on three cutting-edge network intrusion detection datasets (RoEduNet-SIMARGL2021, NSL-KDD, and CICIDS-2017). Future research on this topic could benefit from examining the effectiveness of our XAI evaluation framework on additional network intrusion detection benchmark datasets, such as vulnerability-based datasets~\cite{dong2019towards}, UMass~\cite{UMASS_IDS}, and UNSW-NB15~\cite{moustafa2015unsw}.

\textbf{Dataset Analysis:} The experiments allow us to extract a few interesting conclusions about the datasets used. First, Table~\ref{tab: Summary Metrics Table} shows that LRP has better  performance but the performance is still dataset-dependent to some extent. For example, IG is the best XAI method for CICIDS-2017, while DeepLift and IG are the best for RoEduNet-SIMARGL2021. One characteristic of the RoEduNet-SIMARGL2021 dataset is the lower number of classes, lower number of features, and a humongous number of samples. Hence, there could exist a correlation between DeepLift performing better when these characteristics are present, since it marked best four times out of seven, while it marked one time out of seven for the NSL-KDD and CICIDS-2017. In this case, both datasets have a higher number of features and classes. Lastly, LRP and IG performances are not deeply different dataset-wise, in other words, they still mark at least three out of seven marks for all cases indicating quality of versatility for different scenarios.

\textbf{Reliability of Current White-box XAI Methods:} Although LRP, IG, and DeepLift can be used for auditors when gathering information and understanding network traffic logs for network IDS. Our work shows that the performance of  LRP, IG, and DeepLiftwould need to be improved to be used in real-world IDS. In particular, our work shows that it is desirable to enhance  LRP, IG, and DeepLift to be more robust against adversarial attacks. Furthermore, it shows the need to validate the completeness of explanations from  LRP, IG, and DeepLift before deploying them in reality. 

\textbf{Development of White-box XAI Methods:} The potential for an improved White-box XAI tool to assist network security analysts in saving time and enhancing their assertiveness and efficiency is demonstrated by this work. This study demonstrates that such a potent XAI tool ought to eventually be successful in each of the six measures this work evaluates.  Our robustness test also revealed gaps in the white-box XAI explanations. Specifically, it recommends strengthening the XAI techniques' resilience to hinder hostile assaults. In addition, the efficiency experiment demonstrates 
great potential for the white-box XAI. Nonetheless, it is crucial to be attentive to its limitations, for example, IG and LRP can show memory and time complexity when the demand is high, as seen in the Completeness experiment. Also, DeepLIFT displayed limited performance as the number of samples analyzed increased, which could be detrimental to time-sensitive operations. 
Similar to this, criteria for descriptive accuracy, sparsity, and stability showed appropriate consistency of XAI outcomes. All things considered, these findings and recommendations can help get us closer to our ultimate objective of creating a strong white-box XAI solution for network intrusion detection. 

\textbf{Incorporation of other White-box methods in Future XAI Evaluations:} This work evaluated LRP, IG, and DeepLift in the context of IDS. However, it is crucial to keep exploring other white-box XAI methods, such as PatternNet, PatternAttribution~\cite{pattern}, DeConvNet~\cite{devcon}, GuidedBackProp~\cite{GBP}, CAM\cite{cam}, GradCAM\cite{gradcam}, and GradCAM++\cite{gradcam+} for extended insights in the field.

\textbf{Time Performance Variations Among White-Box XAI Methods:} 
Our investigation shows that LRP and IG perform significantly differently in terms of time efficiency as compared to DeepLift. When the sample size grows from one to ten thousand, DeepLift for all three datasets is slower; it goes from 0.1 to 1380.1 seconds, while the maximum figure for LRP and IG was 14.1 seconds (i.e., 100x lower). Additionally, DeepLift takes longer when there are more features. For instance, CICIDS-2017 (1323.7 seconds) contains more features than the dataset (651.6 features). In contrast to RoEduNet-SIMARGL2021 (14.1 seconds), which performs better for CICIDS-2017 (0.8 seconds), IG and LRP appear to exhibit the opposite tendency.

\section{Conclusion}\label{sec: conclusion}

An IDS works as a prevention or barrier watching the network against potential threats, and with the aid of AI, it has become even more reliable, autonomous, and ubiquitous. Nonetheless, the trade-off of the enhanced prediction power is the higher complexity in explaining the reasoning behind AI tools, which creates a gap in explaining tools, namely XAI (eXplainable Artificial Intelligence). This work proposes an end-to-end framework (i.e., from loading the datasets to applying evaluation metrics) for analyzing and explaining the white-box XAI methods, specifically white-box XAI methods in the field of network intrusion detection. In summary, three invaluable datasets (i.e., RoEduNet-SIMARGL2021, CICIDS-2017, and NSL-KDD) were analyzed using DNN models, three powerful white-box XAI methods (i.e., Integrated Gradients (IG), Layer-wise Relevance Propagation (LRP), and DeepLift), and six evaluation metrics (i.e., Descriptive Accuracy, Sparsity, Stability, Efficiency, Robustness, and Completeness). 

The six metrics evaluated include Descriptive accuracy, which assesses the balance between XAI explainability and intrusion detection accuracy. Sparsity measures the correlation between various intrusion features identified by the XAI method. Stability checks the consistency of the XAI method in generating explanations for traffic instances. Efficiency evaluates the applicability of the XAI method in real-world systems.  Robustness tests the XAI method's ability to produce consistent explanations under adversarial attacks. Completeness verifies the XAI method's capacity to provide accurate explanations for all possible network traffic samples, including edge cases.

In addition, white-box XAI methods have an advantage over black-box methods as they inherently offer model access, making them more comprehensive and effective for real-world applications. This accessibility facilitates troubleshooting and ensures valid explanations. Our experiments demonstrated that white-box XAI methods are more resilient to perturbations and adversarial attacks. Additionally, the results indicated that white-box methods matched or surpassed black-box methods in all metrics except Stability. Hence, it is advised to use white-box XAI methods over black-box XAI methods when possible.

Lastly, the framework was extensively evaluated and validated under the experiment conducted over the three highly-regarded datasets, RoEduNet-SIMARGL2021, CICIDS-2017, and NSL-KDD. Moreover, the code used in this comprehensive study is open-access and available to the community to serve as a means to progress the white-box XAI applied to network intrusion detection, representing an important step in this direction.







\section*{Declarations}

\subsection*{Ethical Approval and Consent to participate}

Not applicable.

\subsection*{Availability of supporting data}

The datasets generated and/or analysed during the current study are available in the GitHUB repository, \url{https://github.com/ogarreche/XAI_Whitebox}.

\subsection*{Competing interests/Authors' contributions}

The authors declare that they have no competing interests.

\subsection*{Funding}

This work is supported in part by AnalytixIN, Enhanced Mentoring Program with Opportunities for Ways to Excel in Research (EMPOWER), and 1st Year Research Immersion Program (1RIP) grants from the office of the Vice Chancellor for Research at
Indiana University-Purdue University Indianapolis.

\subsection*{Author contribution}

Conceptualization: O.A. and M.A.; Data curation: O.A.; Formal analysis: O.A.; Funding acquisition: M.A.; Investigation: O.A.; Methodology: O.A.; Project administration: M.A.; Resources: M.A.; Software: O.A.; Supervision: M.A.; Validation: O.A.; Visualization: O.A.; Writing – original draft: O.A.; Writing - review \& editing: O.A. and M.A.

\subsection*{Acknowledgements}

Not applicable.




\noindent

\bigskip





\bibliography{sn-bibliography}

\begin{appendices}

\section{Hyper-parameters}\label{app:ai_models_hyperparams}

This section displays the hyper-parameters of the Tensor-flow-based DNN models for each dataset.\vspace{-2mm}

\subsection{Hyper-parameters Details}

\textbf{RoEduNet-SIMARGL2021:} For this dataset the architecture consists of the input layer, in which the number of nodes is dependable on how many features are used for the experiment. It consists of three dense layers, they have 128, 64, and 32 nodes respectively, that use relu activation. The output layer has three nodes for its classes. The optimizer is ADAM with a learning rate of 0.001. Plus, the loss is set to sparse\_categorical\_crossentropy, Batch size is the size length of X\_train, epochs to 1000 with early stopping (patience set to 10), and metrics to accuracy. 

\textbf{CICIDS-2017:} For this dataset the architecture consists of the input layer, in which the number of nodes is dependable on how many features are used for the experiment. It consists of one dense layer with seven nodes using relu activation and one output layer of seven classes. The optimizer is ADAM with a learning rate of 0.01. Plus, the loss is set to sparse\_categorical\_crossentropy, Batch size is the size length of X\_train, epochs similar to prior dataset. 

\textbf{NSL-KDD:} For this dataset, the architecture consists of the input layer, in which the number of nodes is dependable on how many features are used for the experiment. It consists of one dense layer with five nodes using relu activation and one output layer of five classes. The optimizer is ADAM, with a learning rate of 0.1. Plus, the loss is set to sparse\_categorical\_crossentropy, Batch size is the size length of X\_train, epochs similar to other datasets. 

\subsection{Hyper-parameter Rationale} 

The models were extensively trained and experimented with different parameters and architectures before they were applied to the study. In addition, the metrics used to direct the efforts were well established in the area (i.e.,  accuracy (Acc), precision (Prec), recall (Rec), F1-score (F1)). For instance, we first conditioned the number of nodes of the input layer to be the same as the features considered and the output layer to be a number of classes. Note that because of incompatibility with iNNvestigate, we could not use the `softmax' activation. Next, we experimented a few times with the number of nodes in a single inner layer until the number 128. Then, after picking the best result, more layers were added in the same fashion. The best parameters were kept for this phase.

\section{List of Top Intrusion Features}\label{app:feat}
We show the list of top intrusion features for both IG, LRP, and DeepLift XAI methods for different datasets in Table~\ref{tab:XAI_topfeatures}.

\begin{table*}
    \centering
    \caption{The best features for the RoEduNet-SIMARGL2021, CICIDS-2017, and NSL-KDD datasets considering IG, DeepLift, and LRP.}
    \resizebox{\linewidth}{!}{
    \begin{tabular}{c|c|c|c}
    \hline
         \textbf{RoEduNet-SIMARGL2021}&  \textbf{DeepLift}&  \textbf{IG}& \textbf{LRP}\\
         \hline
         1&  TCP\_WIN\_MIN\_IN
&  TOTAL\_FLOWS\_EXP
& TOTAL\_FLOWS\_EXP
\\
         2&  TCP\_WIN\_MIN\_OUT
&  LAST\_SWITCHED
& LAST\_SWITCHED
\\
         3&  TCP\_WIN\_SCALE\_OUT
&  TCP\_WIN\_MAX\_IN
& TCP\_WIN\_MIN\_IN
\\
         4&  FLOW\_DURATION\_MILLISECONDS
&  TCP\_WIN\_MIN\_IN
& TCP\_WIN\_MAX\_IN
\\
         5&  TCP\_WIN\_MAX\_IN
&  FIRST\_SWITCHED
& FIRST\_SWITCHED
\\
         6&  TCP\_WIN\_SCALE\_IN
&  L4\_SRC\_PORT
& TCP\_WIN\_SCALE\_IN
\\
         7&  TCP\_WIN\_MAX\_OUT
&  TCP\_WIN\_SCALE\_IN
& L4\_SRC\_PORT
\\
         8&  FLOW\_ID
&  FLOW\_ID
& L4\_DST\_PORT
\\
         9&  L4\_SRC\_PORT
&  L4\_DST\_PORT
& FLOW\_ID
\\
 10& TOTAL\_FLOWS\_EXP
& PROTOCOL
&PROTOCOL
\\
\hline
 \textbf{CICIDS-2017}&  \textbf{DeepLift}&  \textbf{IG}& \textbf{LRP}\\
         \hline
         1&  PSH Flag Count
&  min\_seg\_size\_forward
& min\_seg\_size\_forward
\\
         2&  ACK Flag Count
&  PSH Flag Count
& PSH Flag Count
\\
         3&  Bwd Packet Length Mean
&  ACK Flag Count
& ACK Flag Count
\\
         4&  Init\_Win\_bytes\_forward
&  Init\_Win\_bytes\_forward
& Init\_Win\_bytes\_forward
\\
         5&  Bwd Packet Length Std
&  Destination Port
& Destination Port
\\
         6&  Fwd Packets/s
&  Bwd Packet Length Mean
& Bwd Packet Length Mean
\\
         7&  Packet Length Std
&  Fwd Packets/s
& Fwd Packets/s
\\
         8&  Packet Length Variance
&  Init\_Win\_bytes\_backward
& Init\_Win\_bytes\_backward
\\
         9&  Destination Port
&  Packet Length Mean
& Packet Length Mean
\\
 10& Avg Bwd Segment Size
& Packet Length Std
& Packet Length Std
\\
 \hline
 \textbf{NSL-KDD}&  \textbf{DeepLift}&  \textbf{IG}& \textbf{LRP}
\\
         \hline
         1&  serror\_rate&  same\_srv\_rate& flag\_S0
\\
         2&  flag\_S0&  service\_http& flag\_SF
\\
         3&  dst\_host\_srv\_serror\_rate&  flag\_SF& srv\_serror\_rate
\\
         4&  flag\_SF&  dst\_host\_count& dst\_host\_count
\\
         5&  srv\_serror\_rate&  service\_private& dst\_host\_srv\_count
\\
         6&  dst\_host\_serror\_rate&  logged\_in& dst\_host\_srv\_serror\_rate 
\\
         7&  dst\_host\_srv\_count&  count& serror\_rate
\\
         8&  dst\_host\_same\_srv\_rate&  dst\_host\_serror\_rate& service\_http
\\
         9&  service\_http&  flag\_S0& count
\\
 10& dst\_host\_count& Protocol\_type\_tcp& dst\_host\_rerror\_rate
\\
 \hline

 \end{tabular}
 }
    \label{tab:XAI_topfeatures}
\end{table*}

\end{appendices}


\end{document}